\documentclass[preprintnumbers,amsmath,amssymb,prd, notitlepage,nofootinbib,twocolumn, superscriptaddress]{revtex4-1}

\usepackage{graphicx}
\usepackage{dcolumn}
\usepackage{bm}
\usepackage{subfigure}
\usepackage{wasysym}
\usepackage{verbatim}
\usepackage{color}
\usepackage{mathtools}
\usepackage{hyperref}
\usepackage{soul}

\usepackage{slashed}
\usepackage{amsmath}
\usepackage{multirow}
\usepackage{booktabs}  
\usepackage{natbib}
\usepackage{float}
\usepackage{afterpage}
\usepackage{dcolumn}
\usepackage{bm}
\usepackage{amssymb,amsmath}  
\usepackage{lipsum}
\usepackage{setspace}
\usepackage{etoolbox}

\def\beq{\begin{equation}}
\def\eeq{\end{equation}}
\def\L{{\bf L}}

\newcommand{\pot}{\phi}               

\newcommand{\tl}{\tilde{T}}                 
\newcommand{\tu}{{T}}         

\newcommand{\begm}{\begin{pmatrix}}
\newcommand{\enm}{\end{pmatrix}}

\renewcommand{\eqref}[1] {equation $($\ref{#1}$)$}
\newcommand{\be}{\begin{equation}}
\newcommand{\ee}{\end{equation}}
\newcommand{\ba}{\begin{eqnarray}}
\newcommand{\ea}{\end{eqnarray}}

\def\bL{\bmm{L}}
\newcommand{\bmm}[1]{{\mathbf{#1}}}

\newcommand{\bx}{{\bf x}}
\newcommand{\br}{{\bf r}}
\newcommand{\bk}{{\bf k}}

\def\tuF{\theta_{\mathrm{FWHM}}}

\def\mnras{Mon. Not. R. Astron. Soc}

\begin{document}
\title{Improving Small-Scale CMB Lensing Reconstruction}
\author{Boryana Hadzhiyska}
\affiliation{Harvard-Smithsonian Center for Astrophysics, 60 Garden St., Cambridge, MA 02138, USA}
\affiliation{Department of Applied Mathematics and Theoretical Physics, University of Cambridge, Wilberforce Road, Cambridge CB1 2AD, United Kingdom}
\author{Blake D. Sherwin}
\affiliation{Department of Applied Mathematics and Theoretical Physics, University of Cambridge, Wilberforce Road, Cambridge CB1 2AD, United Kingdom}
\affiliation{Kavli Institute for Cosmology Cambridge, University of Cambridge, Wilberforce Road, Cambridge CB1 2AD, United Kingdom}
\author{Mathew Madhavacheril}
\affiliation{Department of Astrophysical Sciences, Princeton University, Princeton, NJ 08544, USA}
\author{Simone Ferraro}
\affiliation{Lawrence Berkeley National Laboratory, One Cyclotron Road, Berkeley, CA 94720, USA}
\affiliation{Berkeley Center for Cosmological Physics, University of California, Berkeley, CA 94720, USA}

\begin{abstract}
Over the past decade, the gravitational lensing of the Cosmic Microwave Background (CMB) has become a powerful tool for probing the matter distribution in the Universe. The standard technique used to reconstruct the CMB lensing signal
 employs the quadratic estimator (QE) method, which has recently been shown to be suboptimal for lensing
 measurements on very small scales in temperature and polarization data. We implement a 
simple, more optimal method for the small-scale regime, 
which involves taking the direct inverse of the
 background gradient. We derive new techniques to make continuous maps of lensing using this ``Gradient-Inversion'' (GI) method and validate
 our method with simulated data, finding good agreement with predictions. For idealized simulations of lensing cross- and autospectra that neglect foregrounds, we demonstrate that our method performs significantly better than previous quadratic estimator
methods in temperature; at $L=5000-9000$, it reduces errors on the lensing auto-power spectrum 
by a factor of $\sim 4$ for both idealized CMB-S4 and Simons Observatory-like experiments and by a factor of $\sim 2.6$ for cross-correlations of CMB-S4-like lensing reconstruction and the true lensing field. We caution that the level of the neglected small-scale foreground power, while low in polarization, is very high in temperature; though we briefly outline foreground mitigation methods, further work on this topic is required. Nevertheless, our results show the future potential for improved small-scale CMB lensing 
 measurements, which could provide stronger constraints on
 cosmological parameters and astrophysics at high redshifts.
\end{abstract}
\maketitle

\section{Introduction}
Building on progress in CMB experimentation and theory, the gravitational lensing of the cosmic microwave background (CMB) has recently emerged as a powerful cosmological probe. The CMB lensing signal arises from the deflection of CMB photons as they pass through the matter distribution between the surface of last scattering and us; this lensing deflection induces subtle, non-Gaussian correlations
of the CMB anisotropies in both temperature and polarization. 
Reconstructing and analyzing a lensing map from CMB data can provide a wealth of information on the sum of the neutrino masses, the equation of state of dark energy, the properties of inflation and the early universe (via delensing), and astrophysics at high redshift (through cross-correlations) \cite{2006PhRvD..74l3002S,2018arXiv180706210P}. Rapid progress in the extraction of the lensing signal has been made only recently with the first detections in cross- and auto-correlation in 2007 \cite{Smith:2007rg} 
and 2011 respectively  \cite{2011PhRvL.107b1301D}, followed by significantly improved measurements by the SPT, \textit{Planck}, SPTpol and POLARBEAR collaborations \cite{2012ApJ...756..142V,2017ApJ...849..124O,Ade:2015zua,2015ApJ...810...50S,Ade:2013gez}.
While the field has already advanced significantly, CMB lensing science still has great potential, with large increases in the lensing signal-to-noise ratio expected from ongoing and upcoming ground-based CMB experiments such as AdvancedACTPol, SPT-3G, POLARBEAR-II, Simons Observatory and CMB-S4.

Lensing is most commonly reconstructed using the quadratic estimator (QE) proposed by Hu and Okamoto \cite{hu2002mass}. It was shown by 
\cite{hirata2003analyzing,Horowitz:2017iql,carron2017maximum} that in the low-noise
regime in polarization, the QE is suboptimal, as it is only an approximation to the much more computationally expensive
 optimal (maximum likelihood) solution. While it was long thought that the QE was sufficient in temperature, we recently pointed out that this is not the case and that the QE is, in fact, suboptimal on small angular scales in temperature \cite{Horowitz:2017iql}. For small-scale reconstruction from
  temperature, the QE approach is hindered by the fact that
  the precision of the reconstruction is
limited by cosmic variance of the background CMB gradient, whereas a true optimal solution, in principle, should not be.
Furthermore, its errors become highly correlated on small-scales, so that measuring more modes 
does not offer significant improvements when one accounts for the covariance between them properly.
More optimal methods suffer neither of these problems on small scales. We recently showed that on very small scales the so called ``Gradient-Inversion'' (GI) approach, the idea of
which was first proposed by \cite{seljak2000lensing}, can avoid the QE cosmic-variance limit and is close to optimal for reconstruction of these small-scale lenses \cite{Horowitz:2017iql}. However, in our previous work, we only discussed the relevant estimator assuming the approximation of a small, constant-gradient patch of sky, and only analyzed its impact on cluster lensing reconstruction.

In this work, we derive the most general GI lensing algorithm, allowing the reconstruction of a large-area, continuous small-scale lensing map from CMB data; we characterize this estimator's performance for both lensing auto- and cross-spectra. Our paper is structured as follows. In section II we briefly review the QE and its limitations, contrasting it with the GI method proposed by Horowitz et al. \cite{Horowitz:2017iql}. In section III we present a new derivation of the continuous GI estimator for small-scale lensing reconstruction and describe the implementation of our algorithm. Finally, we focus on the signal-to-noise ratio improvement over the standard QE method and discuss systematic error challenges and potential applications of our methods.

\section{Motivation}

The lensing of the CMB can be described 
by a remapping of the unlensed CMB tempearture field $\tu$ to give a lensed field $\tl$:
\be
\tl(\bx) =\tu(\bx+\nabla \pot) \approx  \tu(\bx)+\nabla T \cdot \nabla \pot ,
\label{eq:approx}
\ee
where we have expanded to first order in the lensing potential $\pot$.
Although this first-order approximation can be poor at the field level, it works quite well on small
 enough scales, where the primary temperature fluctuations are severely
 suppressed by diffusion damping. We can find which scales contribute to the gradient
of the temperature by computing its variance up to some multipole $L_{\rm max}$
\begin{equation}
G_{\rm rms}^2(<L_{\rm max})= \int^{L_{\rm max}}_{0} \frac{d^2 \bL}{(2\pi)^2} \ L^2C_L^{\tu \tu}
\label{eq:G_rms},
\end{equation}
where we are working in the flat sky approximation.
As discussed in
\cite{hu2007cluster,Horowitz:2017iql},
the gradient variance saturates at $L_G \approx 2000$.
 Therefore, if we restrict ourselves to small lenses where $l \gg 2000$, the first-order approximation 
of Eq.~\ref{eq:approx}
should be very good. This is the regime we will concentrate on in this work.  

Neglecting the unlensed temperature in this small-scale regime due to Silk damping,
the gradient inversion (GI) solution for the lensing potential in a small patch 
of sky with a single (constant) background gradient was proposed by \cite{seljak2000lensing}
and extended by \cite{Horowitz:2017iql}:
\begin{equation}
\hat{\pot}_{\rm GI}(\L) = \frac{\tl(\L)}{i \bL \cdot \nabla \tu} \ , 
\end{equation}
where $\nabla \tu$ is the constant gradient of the unlensed
CMB temperature in the small patch of sky considered.
If we recall that $\tilde{T}$ contains instrument noise $n$ as well as signal, i.e. that $\tilde{T} = \tilde{T}^{true}+n$ (where $n$ has noise power spectrum $N_L^{TT}$), we can see that the noise level depends on the local gradient:
\begin{equation}
  N_\bL^{\phi \phi} \sim \frac{N_L^{TT}}{(\bL \cdot \nabla T)^2  } .
\end{equation}
We note that in the first expression for $\hat{\pot}_{\rm GI}(\L)$,
$\tl(\L)$ and $\nabla \tu$  can in principle be measured with
 arbitrary accuracy assuming low enough noise, if we make the approximation that the effects of lensing on the large scale gradient are small, i.e.~that $\nabla T$ can be measured as $\nabla \tilde{ T}\approx \nabla T$.
 
In the small-scale limit, the QE estimator has the following
form:
\begin{equation}
\hat{\pot}_{\rm QE}(\L) \approx  \frac{\tl(\bL)}{L^2}  \frac{ \nabla \tu \cdot \bL}{\frac{1}{2}|\nabla \tu|_{\rm rms}^2}\approx \hat{\pot}_{\rm GI}(\L) \frac{ (\nabla \tu \cdot \bL)^2}{\langle(\nabla \tu \cdot \bL)^2\rangle}_{\mathrm{CMB}} ,
\end{equation}
 where the notation $\langle ... \rangle_{\mathrm{CMB}}$ refers to an averaging over realizations of the unlensed CMB. In contrast to the GI estimator, the standard QE does something suboptimal in the low-noise limit: rather than ``correctly'' dividing out the gradient, it multiplies by the gradient and divides by the gradient RMS to preserve correct normalization. This implies that the QE is limited by an unnecessary cosmic variance error in the low-noise limit; in addition, the error is correlated across different modes which share a common gradient. 

While, under the approximations outlined previously, the GI method is close to optimal in the small-scale limit, it has thus far only been defined for a constant gradient. In the next section we will discuss the extension to varying gradients.

\section{The continuous gradient inversion estimator}
The standard gradient inversion estimator was derived in the limit of a very small map with a constant gradient. In this section, we will attempt to make our derivation of the estimator form more rigorous and more generally applicable, extending our lensing reconstruction algorithm to large, real maps. Our strategy will be to relate our final target -- Fourier space quantities such as power spectra -- to an intermediate quantity involving real space correlation functions, where we can more easily apply approximations that are naturally made in real space, such as assuming slowly varying local gradients.
\label{sec:theo}
\subsection{Deriving the estimator}
\label{subsec:deriv}
We will begin by deriving a large-map GI estimator in a simplified and suboptimal manner, neglecting instrumental noise and foregrounds and weighting all regions of the map equally. (A better weighting that maximizes signal-to-noise will be derived later.) Let us define the estimator:
\beq
\hat{\phi} (\bx) \equiv  {\tl(\bx)_{\text{high-pass}}}{f (\nabla T(\bx))} ,
\label{eq:ML_est}
\eeq
where $f (\nabla T)$ is a weight function that only depends on $\bx$ via the slowly varying gradient field (which is assumed to be well-known). Initially, our goal will be to derive $f$ such that it gives an unbiased lensing reconstruction.

To begin, we may write
\beq
\hat{\phi} (\bx) =  \sum _j {\tl(\bx)_{\text{high-pass}}}{f( \nabla T(\bx))} P_j(\bx) ,
\eeq
where we have introduced ``large-pixel'' window functions $P_j$, which are much larger than the lensing scales of interest, but are sufficiently small that we can assume the temperature gradient is constant within the large pixels. The pixel windows $P_j$ are zero outside the $j$-th pixel and equal to one inside. Since the map is high-pass filtered, we can assume that $\tl(\bx)_{\text{high-pass}} \approx \nabla \phi \cdot \nabla T$ so that 
\beq
\hat{\phi} (\bx) =  \sum _j { \nabla \phi(\bx) \cdot \nabla T(\bx)}{f(\nabla T(\bx))} P_j(\bx) .
\eeq
We can similarly write the true $\phi$ field in terms of the large pixelization:
\beq
\phi(\bx)= \sum_j \phi(\bx) P_j(\bx) .
\eeq

To correctly reconstruct the field $\phi$ using an estimator $\hat \phi$, our requirement is that
\beq
C_\bL^{\phi \hat \phi} = C^{\phi \phi}_\bL .
\eeq

We can write this power spectrum in terms of the correlation function

\beq
C_\bL^{\phi \hat \phi} = \int d^2 r  e^{i \bL \cdot \br}\langle  \phi (\bx +\mathbf{r}) \hat \phi(\bx) \rangle_{\mathrm{all}},
\label{eq:powerSpecDefinition}
\eeq
where $\langle ... \rangle_{\mathrm{all}}$ indicates both i) an ensemble average over realizations of small-scale CMB, lensing and noise and ii) a spatial average over $\bx$ within a very large region with many independent gradients. However, throughout this analysis, we will assume the gradients $\nabla T(\bx)$ to be {fixed} and, as stated previously, assume them to be well known. In some cases it will be helpful to average over small-scale fields, denoted $\langle ...\rangle_{\mathrm{s}}$, and $\bx$, marked $\langle ...\rangle_{\bx}$, in separate steps. (This implies $\langle ... \rangle_{\mathrm{all}}=\langle \langle ...\rangle_{\mathrm{s}} \rangle_{\bx}$.)

Now we may evaluate the real space correlation function, which will allow us to make a number of relevant approximations. Averaging only over small-scale fields in a first step, we obtain

\begin{eqnarray}
\langle  \phi (\bx +\mathbf{r}) \hat \phi(\bx) \rangle_{\mathrm{s}} =&&
\sum _{j,k} P_j(\bx)P_k(\bx+\br) \times 
\\ &&~~ \langle { \phi(\bx+\br) \nabla \phi(\bx) \cdot \nabla T(\bx)}{f( \nabla T(\bx))}\nonumber \rangle_{\mathrm{s}}.
\end{eqnarray}

We may assume that, since we have high pass filtered to remove long wavelength modes in our estimator (and are, in any case, only considering lensing on small scales, which approximately corresponds to very small $\br$ correlations), only correlations within the same pixel $j = k$ are significantly non-zero, and we will neglect all others. Therefore only the $j=k$ terms contribute to the sum. In addition, we may assume that within one pixel $j$, the temperature gradient is not a varying field, but approximately takes on a constant value $\nabla T_j$. This gives us:

\beq
\langle  \phi (\bx +\mathbf{r}) \hat \phi(\bx) \rangle_{\mathrm{s}} \approx \sum _{j} P_j^2(\bx) {f( \nabla T_j)}{ \langle \phi(\bx+\br) \nabla \phi(\bx)\rangle_{\mathrm{s}} \cdot \nabla T_j} .
\eeq
We now write the potential-potential gradient correlation function in Fourier space to give:
\beq
\langle  \phi (\bx +\mathbf{r}) \hat \phi(\bx) \rangle_\mathrm{s} \approx \sum _{j} P_j^2(\bx)  \int \frac{d^2 k}{(2 \pi)^2}  e^{-i \bk \cdot \br} C^{\phi \phi}_\bk {f( \nabla T_j)}{ i\bk \cdot \nabla T_j} .
\label{eq:intermediateCorr}
\eeq
We can now connect this expression to the definition of the power spectrum in terms of the correlation function. For this purpose, we average equation (\ref{eq:intermediateCorr}) over $\bx$, i.e. add $\langle... \rangle_{\bx}$, to
give
\begin{eqnarray}
&&\langle  \phi (\bx +\mathbf{r}) \hat \phi(\bx) \rangle_\mathrm{all} \approx \\&& \sum _{j} \langle  P_j^2(\bx) \rangle_\bx  \int \frac{d^2 k}{(2 \pi)^2}  e^{-i \bk \cdot \br} C^{\phi \phi}_\bk {f( \nabla T_j)} ~{ i\bk \cdot \nabla T_j}\nonumber
\label{eq:intermediateCorr2}
\end{eqnarray}
and insert this two point correlation function into the power spectrum expression of Eq. (\ref{eq:powerSpecDefinition}), noting also that the mean value for $P^2_j$ over $\bx$ is simply $1/N_{\rm pix}$, i.e. one over the number of pixels. We thus obtain

\begin{eqnarray}
&&C_\bL^{\phi \hat \phi} =\\&& \frac{1}{N_{\rm pix}}\sum _{j} \int \frac{d^2 k}{(2 \pi)^2} \int d^2 r  e^{i \bL \cdot \br} e^{-i \bk \cdot \br} C^{\phi \phi}_\bk {f( \nabla T_j)} ~{ i\bk \cdot \nabla T_j}\nonumber .
\end{eqnarray}

Performing the $\br$ integral, we obtain $(2\pi)^2 \delta^{(D)}(\bL-\bk)$; this allows us to perform the $k$ integral to give:

\beq
C_\bL^{\phi \hat \phi} =  \frac{1}{N_{\rm pix}} \sum _{j} C^{\phi \phi}_\bL{f( \nabla T_j)}~ { i\bL \cdot \nabla T_j} .
\eeq
In other words, our cross-correlation measurement is equal to the true cross-correlation in each large pixel, weighted by ${f( \nabla T_j)} ({ i\bL \cdot \nabla T_j}) $ and averaged over all pixels. It can clearly be seen that if we set ${f( \nabla T_j)} = \frac{1}{i \bL \cdot \nabla T_j}$ we obtain
\beq
C_\bL^{\phi \hat \phi} = C^{\phi \phi}_\bL ,
\label{eq:qL}
\eeq
as required, with each large-pixel region contributing equally to the cross-spectrum. 

We note that for the correct reconstruction of each mode of the lensing $\phi_\bL$, we require a different, specific weight function $f^\bL \equiv \frac{1}{i \bL \cdot \nabla T_j}$. We similarly denote the estimator that uses this particular weight function as $\hat \phi^\bL (\bx)$; this estimator is correctly normalized to recover the mode $\bL$; i.e., if we Fourier Transform $\hat \phi^{\bL}(\bx)$, then the mode ${\hat{\phi}}^{\bL}(\bL) \equiv {\hat{\phi}}^{\bL}_\bL $ is correctly reconstructed. Here we have the understanding that a superscripted $^\bL$ implies that a spatial weighting $f^\bL$ is being applied with the purpose of reconstructing $\phi$ at a fixed target wavenumber $\bL$. This is to be distinguished from a subscript $_\bL$, which is the true Fourier conjugate variable of the coordinate $\bx$ describing the real space map. The two variables will be equal if the correct spatial weighting has been applied in order to reconstruct each mode, as will be shown below to be the case in all realistic applications.

In the previously derived simple estimator, all pixels of the map make equal contributions to the final lensing power spectrum. However, a more optimal estimator should up-weight pixels with a large gradient, as these have a larger signal $\nabla T \cdot \nabla \phi$. We will now derive such a more optimally weighted estimator.

Our derivation proceeds similarly to that presented previously. The 
small-scale temperature is now assumed to be described by
$\tl(\bx)_{\text{high-pass}}=\nabla \phi \cdot \nabla T +n$, where noise $n$ has a 
power spectrum $N_L^{TT}$. We again consider the estimator
\beq
{\hat \phi^\bL} (\bx) \equiv  \tl(\bx)_{\text{high-pass}} f^{\bL}(\nabla T) .
\eeq
As before, we obtain for the cross-correlation of the reconstruction with the input:
\beq
C_\bL^{\phi \hat \phi^\bL} =   \frac{1}{N_{\rm pix}} \sum _{j} C^{\phi \phi}_\bL \times (i\bL \cdot \nabla T_j) f^{\bL}(\nabla T_j) .
\label{eq:crossDef}
\eeq
Similarly, we can obtain an expression for the reconstruction power spectrum
\beq
C_\bL^{\hat \phi^\bL  \hat \phi^\bL} =  \frac{1}{N_{\rm pix}} \sum _{j}  [-C^{\phi \phi}_\bL \times  (i\bL \cdot \nabla T_j)^2 +N_L^{TT} ](f^{\bL}(\nabla T_j))^2 .
\label{eq:autoDef}
\eeq
Given these cross- and auto-spectra, we can find the weight function $f^\bL$ that minimizes the variance $\mathrm{Var}(\hat \phi_\bL)$ of our new estimator, which we define as:
\begin{eqnarray}
&& \langle (\hat \phi_\bL - \phi_\bL)(\hat \phi_{\bL'} - \phi_{\bL'})^*\rangle_{\mathrm{all}}\equiv \mathrm{Var}(\hat \phi_\bL)(2 \pi)^2 \delta^{(D)}(\bL-\bL')  \nonumber \\&&~~~~=  (C_\bL^{\hat \phi^\bL  \hat \phi^\bL} - 2 C_\bL^{\phi \hat \phi^\bL} + C_\bL^{\phi \phi})(2 \pi)^2 \delta^{(D)}(\bL-\bL') .
\end{eqnarray}
We must minimize the reconstruction variance $\mathrm{Var}(\hat \phi_\bL)$ subject to the normalization constraint that $C_\bL^{\phi \hat \phi^\bL} = C^{\phi \phi}_\bL$, which is equivalent to 
\beq
A_\bL[f^\bL] \equiv   \frac{1}{N_{\rm pix}} \sum _{j} (i\bL \cdot \nabla T_j) f^{\bL}(\nabla T_j)  = 1.
\eeq
We can perform this minimization using a Lagrange multiplier $\lambda$
\begin{eqnarray}
&&\mathcal L [f^\bL] \equiv\\&&C_\bL^{\hat \phi^\bL  \hat \phi^\bL}[f^\bL] - 2 C_\bL^{\phi \hat \phi^\bL}[f^\bL] + C_\bL^{\phi \phi} - \lambda_\bL(A_\bL[ f^\bL]-1) \nonumber,
\end{eqnarray}
where $C_\bL^{\hat \phi^\bL  \hat \phi^\bL}[f^\bL]$ and $C_\bL^{\phi \hat \phi^\bL}[f^\bL]$ are as in Eqs.~\ref{eq:crossDef} and \ref{eq:autoDef} above.

Setting $\delta \mathcal{L}/\delta f^\bL = 0$ and then imposing the normalization condition $A_\bL[f^\bL]=1$ to solve for $\lambda$, after some algebra we obtain our final weighting for our estimator
\begin{eqnarray}
f^{\bL}(\nabla T_j) =&&\frac{(i\bL \cdot \nabla T_j)}{[-C^{\phi \phi}_\bL (i\bL \cdot \nabla T_j)^2 +N_L^{TT} ]}
\\ &&~~\times \left[\frac{1}{N_{\rm pix}} \sum _{k} \frac{ (i\bL \cdot \nabla T_k)^2}{[-C^{\phi \phi}_\bL  (i\bL \cdot \nabla T_k)^2 +N_L^{TT} ]} \right]^{-1}\nonumber .
\end{eqnarray}
Approximating this as a continuous function, we obtain
\begin{eqnarray}
f^{\bL}(\bx) \approx && \frac{ (i\bL \cdot \nabla T(\bx))}{[-C^{\phi \phi}_\bL (i\bL \cdot \nabla T(\bx))^2 +N_L^{TT} ]}
\\ &&~~\times \left[\left <  \frac{ (i\bL \cdot \nabla T(\bx))^2}{[-C^{\phi \phi}_\bL  (i\bL \cdot \nabla T(\bx))^2 +N_L^{TT} ]} \right >_\bx \right]^{-1}.\nonumber
\end{eqnarray}
Therefore, our final estimator for lensing (factoring out and cancelling several factors) is:
\begin{eqnarray}
{\hat \phi^\bL} (\bx)  \approx &&\frac{ \tl(\bx)_{\text{high-pass}}/(i\bL \cdot \nabla T(\bx))}{[C^{\phi \phi}_\bL  +\frac{N_L^{TT}}{(\bL \cdot \nabla T(\bx))^2 } ]} \times
 \\ &&~~\left[\left < \left[C^{\phi \phi}_\bL  +\frac{N_L^{TT}}{(\bL \cdot \nabla T(\bx))^2}\right]^{-1} \right >_\bx \right]^{-1}.\nonumber
\end{eqnarray}

To optimally recover a mode $\phi_\bL$, we must simply take the Fourier transform (FT) of ${\hat \phi^\bL} (\bx)$ and select the mode with wavenumber $\bL$; i.e.~we must evaluate ${\hat{\pot}}^{\bL}_\bL \equiv \mathrm{FT}[{\hat{\pot}}^{\bL}](\bL)$. The Fourier transform of ${\hat \phi^\bL} (\bx)$, of course, contains a full array of other pixels ${\hat{\pot}}^{\bL}_{\neq \bL}$, but we discard these as our spatial weighting $f^\bL(\bx)$ is not optimal or unbiased for these modes.

In addition, we note that while the estimator contains the unobservable true unlensed gradient, we approximate it by low-pass filtering the observed, lensed map (removing all power above $\ell>2000$) and taking a real-space derivative operation. The tests described later in this work verify that this approximation does not introduce a bias.
Here the second line represents the gradient-dependent normalization that corrects for the weighting of different regions.

Our estimator can be interpreted as an unbiased estimator $ \tl(\bx)_{\text{high-pass}}/(i\bL \cdot \nabla T(\bx))$ multiplied by a weight function that increases the contribution from large-gradient, high signal-to-noise regions. Building on this intuition, our final estimator can be more compactly represented as
\begin{eqnarray}
{\hat \phi^\bL} (\bx)  \approx &&W^\bL ( \bx)\frac{ \tl(\bx)_{\text{high-pass}}}{(i\bL \cdot \nabla T(\bx))} \times \left[\left < W^\bL( \bx) \right >_\bx \right]^{-1},
\label{mainEstimator}
\end{eqnarray}
where we have defined the optimal spatial weighting function as $W^\bL( \bx)\equiv \left[C^{\phi \phi}_\bL  +\frac{N_L^{TT}}{(\bL \cdot \nabla T(\bx))^2 }\right]^{-1} $. However, we will justify in subsequent paragraphs why, to validate our methods using simulations, we use a slightly suboptimal weight $W^\bL( \bx)=\left[\frac{N_L^{TT}}{(\bL \cdot \nabla T(\bx))^2 }\right]^{-1}$. It can be seen that, as expected, $W^\bL$ is simply a spatial window function that upweights high signal-to-noise regions (where the gradient is large) and downweights low signal-to-noise ones.

We note that in the signal-dominated limit, Eq.~(\ref{mainEstimator}) reduces to the naive gradient inversion estimator. 

In the noisy limit, when reconstructing over a large area with many independent gradients, this appears at first glance very similar to the standard quadratic estimator. However, in fact, there is a significant difference to the quadratic estimator, namely how the estimator is normalized. For a single, constant gradient, we have argued previously that the quadratic estimator is suboptimally normalized, making an error of order $\frac{ (\bL \cdot \nabla T 
)^2_{\rm true}}{\langle (\bL \cdot \nabla T)\rangle_{\mathrm{CMB}}^2}$. Naively, spatially averaging over $x$ would drive the GI normalization to the RMS value used in the quadratic estimator. However, the normalization error is reduced only by the square root of the number of independent gradients that are averaged over; since the error on all lensing observables averages down by the same amount, the \emph{relative} impact of this mis-normalization compared to the lensing spectrum error bars is not reduced by spatial averaging. We hence expect the local mis-normalization, arising from the gradient cosmic variance error, to remain a key limiting factor for the quadratic estimator. The GI estimator does not suffer from this limitation.

The GI estimator we have derived and seek to implement, as shown in equation (\ref{mainEstimator}) makes use of a spatial weight (or mask) function $W$; in the optimal case, this should be given by $W^\bL( \bx)=\left[C^{\phi \phi}_\bL  +\frac{N_L^{TT}}{(\bL \cdot \nabla T(\bx))^2 }\right]^{-1}$, effectively a Wiener filter. However, we encountered significant difficulties in implementing this optimal weighting function for low noise levels. A likely source of our implementation problems was the fact that our estimator derivation is only valid for weights which are spatially slowly varying. For very low noise levels, however, the window function varies very rapidly: though $W$ is typically equal to $[C^{\phi \phi}_\bL]^{-1}$, along directions where the gradient is perpendicular to the wavevector $\bL$, $W$ very rapidly takes on a zero weight, leading to sharp edges in the window function. To regularize the weight, we modified our filtering to be a simple inverse noise filter:
\begin{equation}
W^\bL (\bx)\equiv \left[\frac{N_L^{TT}}{(\bL \cdot \nabla T(\bx))^2 }\right]^{-1}.
\label{eq:weight}    
\end{equation}
For noise-dominated modes, this is already (near-) optimal; for signal dominated modes, this leads to moderate loss in the area used, which typically only has a modest impact on signal-to-noise. As is visible from the form of the window function, the theoretical prediction for the noise on our estimator for a particular fixed, local gradient is given by
\begin{equation}
N_\bL^{\hat \pot \hat \pot}(\bx) = \frac{N_L^{TT}}{(\bL \cdot \nabla T(\bx))^2}.
\label{eq:noise_mode}
\end{equation}
The modification to this expression in the case of multiple gradients is 
discussed in more detail in Appendix \ref{app:noise} along with other subtleties
regarding the noise of our estimator.

\subsection{Applying the estimator: summary of the algorithm for lensing reconstruction}
\label{subsec:algo}
In this section we will briefly outline the estimation procedure, Eq.~\ref{mainEstimator}, used for reconstruction of the lensing potential field for a given simulated lensed image using the GI method.

\noindent For each target wavevector $\bL \equiv(L_x,L_y)$ on a discrete FFT grid given by 
spacing $(n_i 2\pi / S_x, n_j 2\pi / S_y)$,
where $S_x$ and $S_y$ are the map dimensions (in this case,
$S_x=S_y=1024\times0.1$ arcmin) and $n_{i,j}$ are integers ranging over the number of pixels, our procedure for obtaining the map and the power spectrum is described by the following steps:
\begin{itemize}
\item[(i)] We obtain the gradient field of the temperature as follows. We first
filter the observed lensed temperature map, i.e. we
remove the small-scale information ($L>2000$), 
where the gradient saturates. We then apply a 
real-space numerical differentiation 
(using second order accurate central differences)
to this filtered field in order to obtain the 
gradient. We denote this gradient of the map as 
$\nabla \tl_{\rm filt}(\bx)$ and
for the subsequent steps of the algorithm,
substitute $\nabla \tu(\bx)$, appearing in the derivations in Section \ref{sec:theo}, with it, as the true unlensed $\nabla \tu(\bx)$ is not observable.

\item[(ii)] Construct the unweighted estimator $\hat \pot^\bL (\bx)_{\rm raw}= \frac{\tl(\bx)_{\text{high-pass}}}{i \bL \cdot \nabla \tl_{\rm filt}(\bx)}$ as defined in Eq.
\ref{eq:ML_est}. At a given position $\bx$, the value of the estimator depends only on the
local gradient $\nabla \tl_{\rm filt}(\bx)$ and the value of the lensed temperature $\tl(\bx)$. (Note that zeros of the gradient are not problematic after the weighting in the next step is applied.)

\item[(iii)] Apply an inverse noise weighting
$W^\bL(\bx) = [N^{\hat \pot \hat \pot}_\bL(\bx)]^{-1}$ to it, i.e.
compute $W^\bL(\bx) \times \hat \pot^\bL (\bx)_{\rm raw}$,
where the window function
is given in Eq. \ref{eq:weight}.
The motivation behind choosing this form for
the filtering function is that it upweights large-gradient, high signal regions; it also downweights 
the modes perpendicular to the gradient direction, as they have infinite
variance.
Thus, we obtain a real-space map of the lensing potential weighted for this specific wavevector.
\item[(iv)] Fourier transform the estimator and keep only the value of the transformed estimator at the target wavenumber $\bL$, i.e. $\hat \pot^\bL (\bx)
\xrightarrow{FT} {\hat{\pot}}^{\bL}(\bL)$ ($={\hat{\pot}}^{\bL}_\bL$). The reason for this step is that the weighting we apply to the reconstructed map 
in the previous step is by construction optimal for the target
wavevector, $\bL$, we chose initially (Eq.~\ref{eq:qL}). To obtain the reconstructed map,
normalize by $1/{\langle W^\bL(\bx)} \rangle$, where averages are over real space (i.e., $\bx$). This step returns the normalized estimator derived in Eq. \ref{mainEstimator}. For each target $(L_x,L_y)$ wavevector pixel, we repeat this procedure to build up the entire reconstructed map in Fourier space. This requires us to perform a full spatial weighting and
Fourier transform of the map for each pixel, which becomes more computationally expensive as the size of the maps is increased (though the algorithm could potentially be sped up by applying the same weighting to blocks of similar $(L_x,L_y)$).

\item[(v)] Compute the 2-dimensional power spectrum. The the cross power spectrum (between the reconstructed and the true fields) is given by
 $C_\bL^{\hat \pot^\bL \pot} = 
 \frac{{({\hat \pot}^{\bL}_\bL)^\ast} \pot_\bL}
 {\langle W^\bL(\bx)   
 \rangle_\bx}$, while auto power spectrum (of the
reconstructed field) can be
obtained as
 $C_\bL^{\hat \pot \hat \pot} =  \frac{{({\hat \pot}^{\bL}_\bL)^\ast} {\hat \pot}^\bL_\bL}{\langle (W^\bL(\bx))^2  \rangle_\bx}$, where for the normalization we have used the standard expressions for calculating pseudo-$C_\ell$ power spectra given a spatially dependent slowly varying window function $W$ \cite{2004MNRAS.349..603E}.
\item[(vi)] Finally, bin the power spectrum with appropriately chosen maximal multipole $L_{\rm max}$ and
number of bins $n_{\rm bins}$. In the subsequent sections, we show the computed
power spectra with $L_{\rm max}=20000$ and $n_{\rm bins}=10$ for an ultra-low noise
experiment and $L_{\rm max}=12000$ and $n_{\rm bins}=12$ for a CMB-S4-like experiment (although much of the signal-to-noise will be at lower multipoles). The small-scale regime is where we expect our method to outperform the QE most significantly.
\end{itemize}

\section{Results and Discussion}
In order to test our method, we apply it to simulations
mimicking data from CMB temperature measurements with different instrumental noise
levels. We simulate the lensing deflection using the following algorithm, 
described in \cite{Nguyen:2017zqu}. First,
we generate a 2-dimensional Gaussian random
field realization from a theoretical CMB unlensed temperature power spectrum
with a pixel size of 0.05 arcmins on a $2048\times 2048$ pixel map.
The CMB maps are remapped (using fifth order spline interpolation to
more accurately apply the
deflection) to produce a lensed temperature map from a Gaussian 
realization of the convergence 
field as predicted from the 
theoretical power spectrum.
The lensed
map is then convolved with a beam of appropriate size and 
instrumental white noise is added. It is then downsampled to 
$1024 \times 1024$ pixel maps in Fourier space in order to
avoid the need to use a pixel window function and to speed up
the lensing reconstruction.
We thus obtain lensed temperature map realizations with simulated instrumental noise
of area $1.707 \times 1.707$ deg$^2$ and a pixel size of 0.1 arcmins.

In this section, we consider three measures of instrumental noise for future experiments:
\begin{itemize}
\item \textbf{Ultra-low (UL) noise experiment:} Futuristic experiment with a noise factor of $\Delta_{\rm T}= 0.1 \ \mu$K-arcmin
and a beam size of $\tuF=0.3$ arcmin.
\item \textbf{CMB-S4-like experiment:} CMB-S4-like experiment
 with a noise factor of $\Delta_{\rm T}=1 \mu$K-arcmin
and a beam size of $\tuF=1 {\text{arcmin}}$.
\item \textbf{SO-like experiment:} 
Simons Observatory (SO)-like experiment
 with a noise level of $\Delta_{\rm T}=6 \mu$K-arcmin
and a beam size of $\tuF=1.4 {\text{arcmin}}$.
\end{itemize} 
The simulations are highly simplified and optimistic in that they contain no Sunyaev-Zel'dovich signal, Cosmic Infrared Background emission or any other extragalactic foreground; we revisit the topic of foreground contamination later. We note that these experiments are highly signal-dominated on large scales in temperature, which means that the
local gradient can be measured with high accuracy.

Throughout this section, we will be comparing
the GI method with the QE one to test whether one
indeed observes an improvement on small scales, as we expect based on our discussion in Section \ref{sec:theo}.
The QE convergence field estimate is obtained 
from the standard TT estimator \cite{Hu_2007} calculated using the modes $\ell_{\rm min} = 1000$ to $\ell_{\rm max} = 20000$. However, we have checked that the QE reconstruction maps give nearly identical results if $\ell_{\rm min}$ is decreased below $1000$ (even to $0$). 
The gradient leg is then low-pass filtered to remove the $L > 2000$
modes.
For the autospectrum calculation, we, furthermore, apply a 
realization-dependent
$N_0$ subtraction to the
QE results (which reduces off-diagonal bandpower correlations) \cite{2013MNRAS.431..609N,PhysRevD.95.123529,2019PhRvD..99b3502N}. Note that
our method is not expected to work on scales with $L$ below $L \sim 4000$, so we apply
a high-pass filter
to the lensed temperature map before 
performing a GI reconstruction on
the simulations. We further note that the statistical power
of the QE is reduced in the small-scale regime 
compared to naive forecasts. This effect is a consequence of the fact that
the errors on the different modes are highly
correlated \cite{Horowitz:2017iql} --
a fact not
accounted for in the theoretical model of its reconstruction noise.

\subsection{Reconstructed Maps}
\label{subsec:recs}
\begin{figure}[!ht]
\centering
    \begin{subfigure}
        \centering
        \includegraphics[width=5cm]{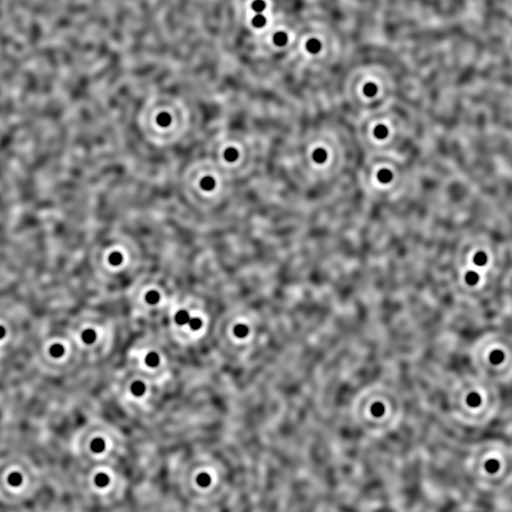}
    \end{subfigure}
    \hfill
    \begin{subfigure}
        \centering
        \includegraphics[width=5cm]{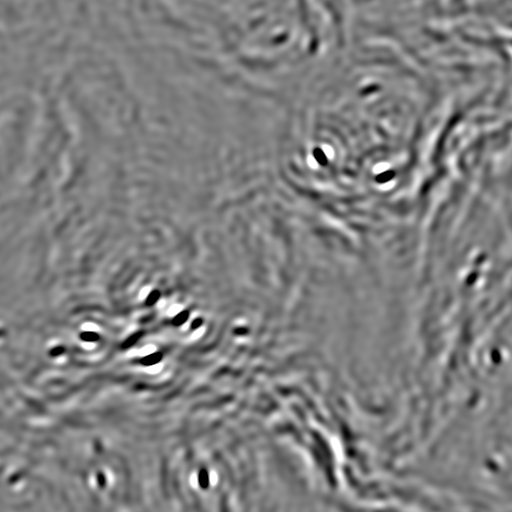}
    \end{subfigure}
    \hfill
    \begin{subfigure}
        \centering
        \includegraphics[width=5cm]{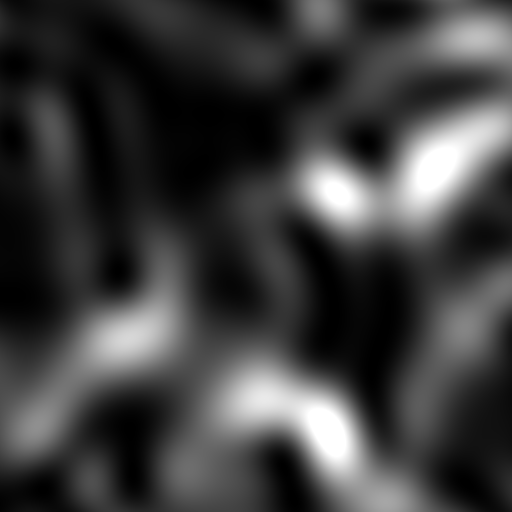}
    \end{subfigure}
    \caption{Map of the input (\textit{upper panel}) and the GI-reconstructed
    convergence field (\textit{center panel}), as well as the magnitude of the temperature gradient
      (\textit{lower panel})
      for a simulated small patch (1.707 $\times$ 1.707 deg$^2$)
of a CMB temperature ultra-low noise experiment with very bright galaxy clusters to better illustrate the reconstruction.
A Wiener filter has been applied to the first two. 
A correlation between the convergence field maps is noticeable by eye as well corresponding to the regions where the magnitude of the gradient is largest.}
\label{fig:recs}
\end{figure}

In Fig. \ref{fig:recs}, we show a comparison between the true
(input)
and the GI-reconstructed lensing convergence fields. We denote the
convergence field by $\kappa(\bx)$ and it is related to the lensing potential
by $\kappa(\bx)=-\frac{1}{2} \nabla^2 \pot(\bx)$.
To aid with visualization of the reconstruction, the input lensing convergence field has been modified with the addition of large, randomly scattered 
massive point sources on top of the CMB lensing convergence map.
We expect that modes with variation
nearly perpendicular to the gradient direction
or patches where the magnitude of the gradient is small
will be reconstructed with very large noise and will, thus, be
downweighted by applying the inverse noise filtering function, $W^\bL(\bx) = [N^{\hat \pot \hat \pot}(\bL,\bx)]^{-1}$,
described in the previous
section. 

Indeed, in regions where the magnitude of the gradient
is large, the
 values of the reconstructed maps appear to be larger as well. This is a consequence of the fact that the inverse noise filter
applied on the reconstructed maps 
downweights patches with small gradients,
and they end up with values
close to zero in the reconstructed plots.
In particular, it is easy to see that some of the artificial point sources
have been recovered quite well by the GI method, while others are
not at all discernible in the reconstruction map. By glancing
at the magnitude of the gradient map, it becomes evident why that is
the case -- the point sources located in positions of the map where
the gradient is large have been correctly identified
by the GI estimator,
while those in regions with small gradient are missing from the
reconstruction. The same qualitative behavior can in fact be observed in the
QE-reconstructed maps as well; however, the weighting of regions with different gradients is expected to be suboptimal for the QE (see end of Subsection \ref{subsec:deriv} for discussion).

\subsection{Estimator Validation and Power Spectra}
\begin{figure}[!ht]
  \centering  \includegraphics[width=0.5\textwidth]{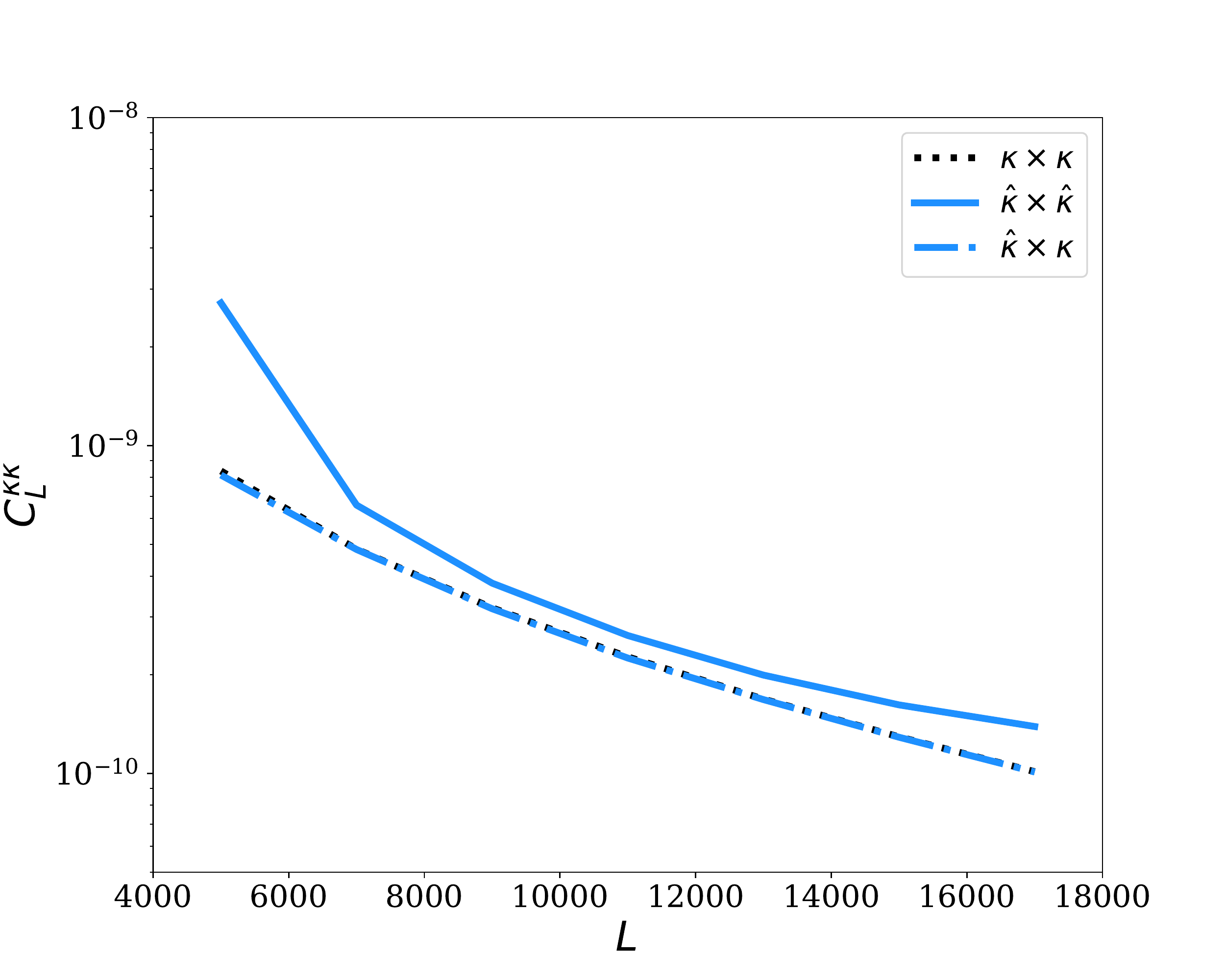}
    \caption{Auto and cross power spectra of the
    reconstructed convergence field for an experiment with ultra-low noise 
      from 360 simulations.
      Here we do not show the $L < 4000$ modes, where we do not expect our 
      method to work well (note our bin width is $\Delta L = 2000$ here).
We see that the cross-correlation with the input is very accurate which
gives credibility to our method. The auto power spectrum remains 
only slightly biased up to large wave vector values, $L \sim 15000$.
The auto power spectrum shown above includes noise bias, which should be
subtracted in a realistic analysis.} 
      \label{fig:cl}
\end{figure}

In Fig. \ref{fig:cl}, we show the auto- and
cross-power spectra (blue solid and dashed-dotted lines, respectively)
obtained through the GI method
from 360 simulations of a small CMB temperature patch
as measured by a CMB-S4-like experiment and
compare them with the auto power spectrum
of the true lensing signal (black dashed curve). The cross power spectrum of the GI-reconstructed lensing matches the true input lensing power to high accuracy, demonstrating that the GI reconstruction algorithm is unbiased and works well for $L>4000$. No bias subtraction is required in this case, which makes this cross-correlation measurement a highly robust validation of our algorithm. 

The auto-power spectrum of the raw reconstructed convergence field, which includes a reconstruction noise bias, is somewhat larger than the signal
on large scales and on small scales $L > 15000$, i.e., it becomes noise dominated. 
Even so, a measurement of the power spectrum is possible
because of the large number of modes.
We note that the reconstruction noise bias present here could be characterized by simulation and subtracted, estimated and removed using data-derived methods (see discussion later), or reduced by cross-correlating different splits of the data. We defer a detailed discussion of robust autospectrum bias subtraction to future work.

\subsection{Simulated Estimator Performance}
\label{subsec:std}

In this subsection, our goal is to characterize the performance of the GI estimator and to compare it with that of the QE estimator, the current standard for reconstructing lensing from CMB temperature.

\begin{figure*}[ht!]
    \centering
    \begin{subfigure}
        \centering
        \includegraphics[width=3.3in]{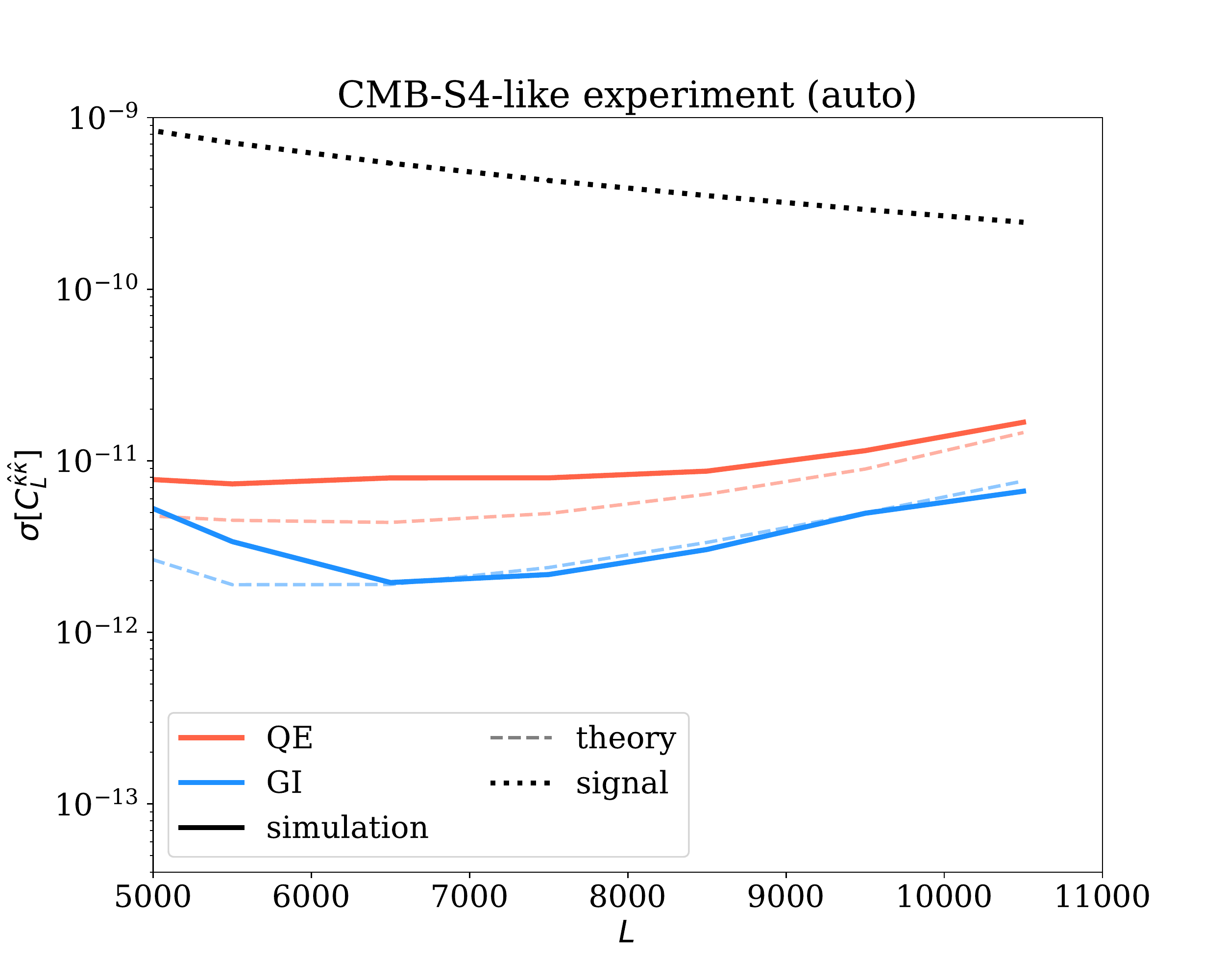}
    \end{subfigure}
    \hfill
    \begin{subfigure}
        \centering
        \includegraphics[width=3.3in]{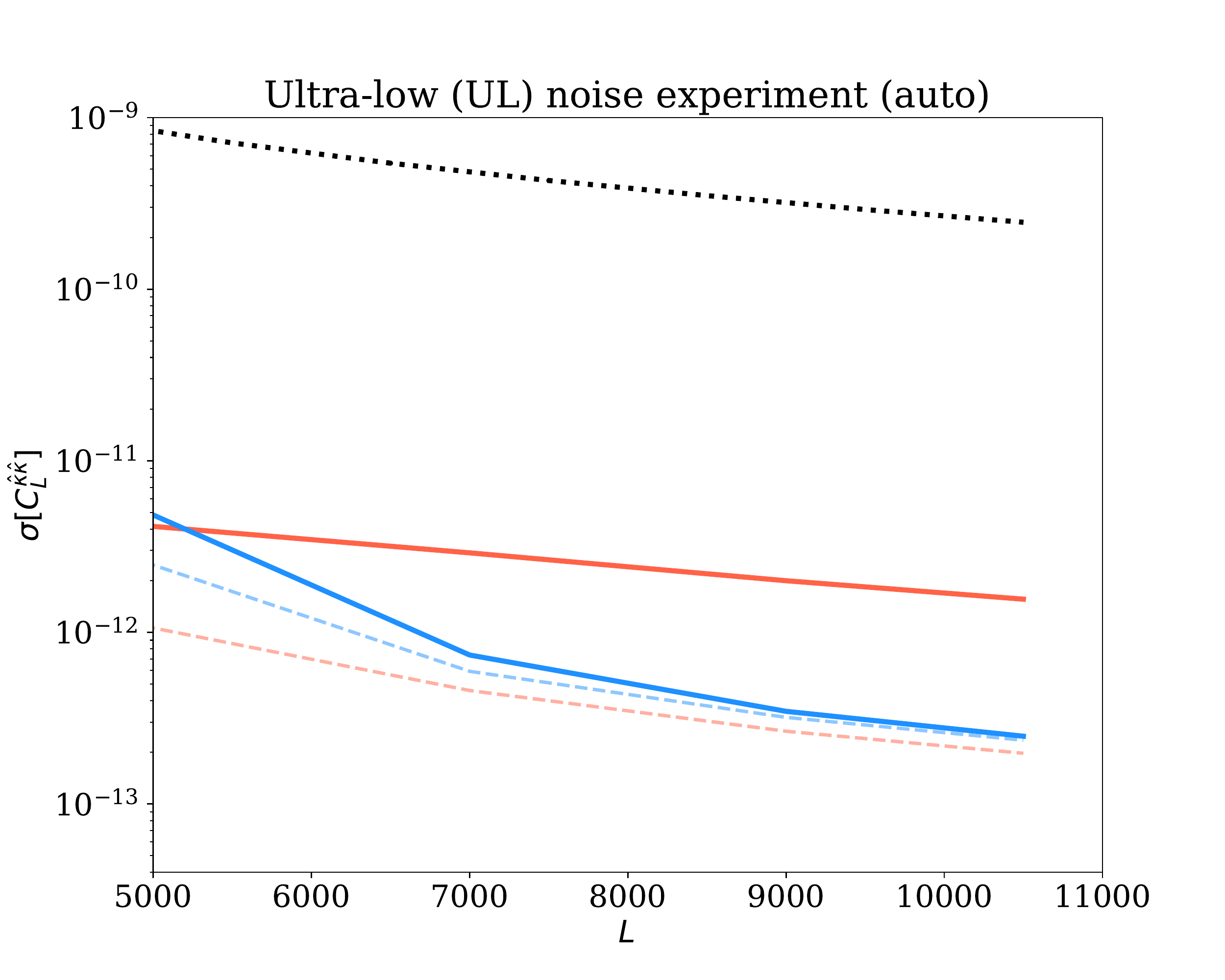}
    \end{subfigure}
    \begin{subfigure}
        \centering
        \includegraphics[width=3.3in]{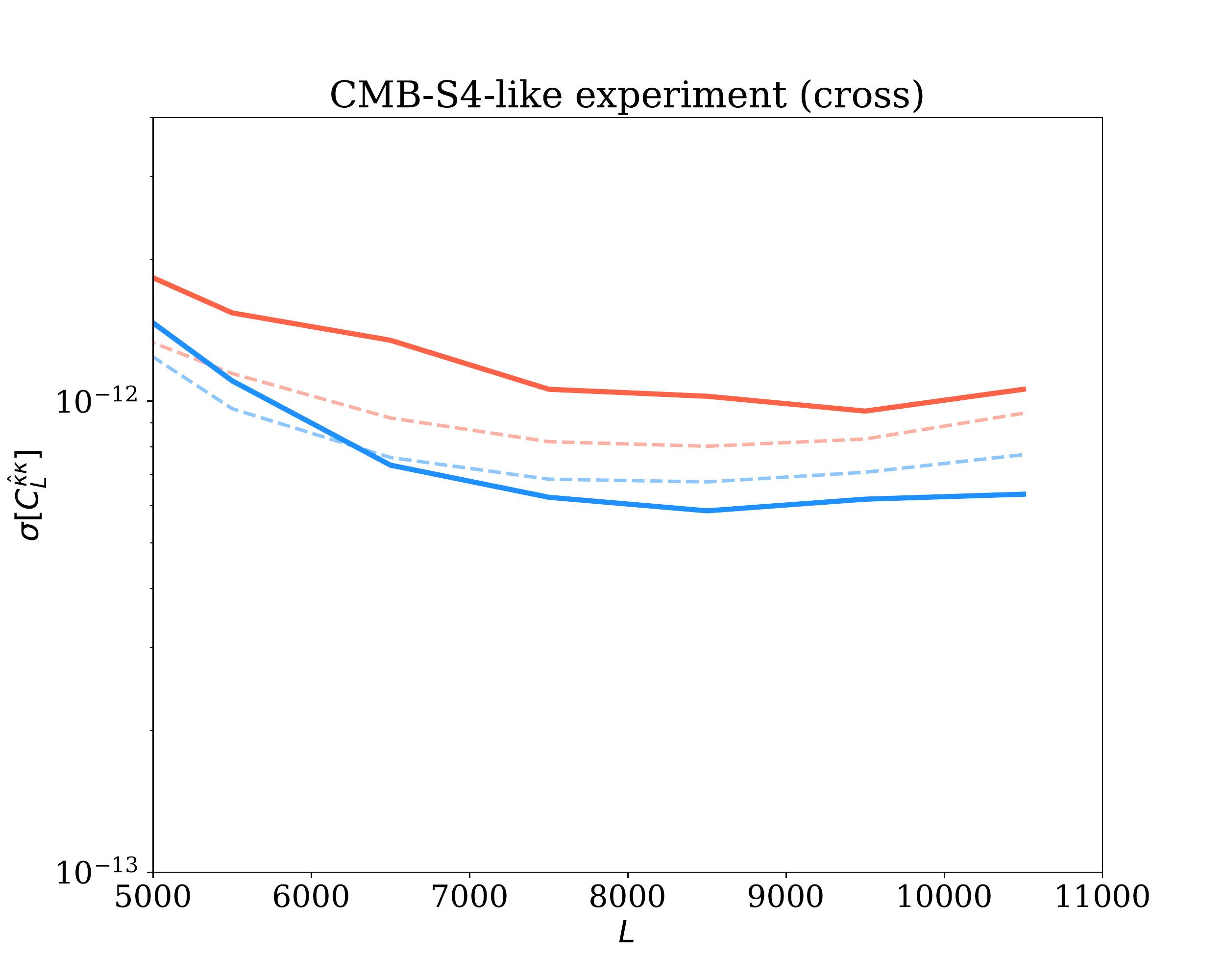}
    \end{subfigure}
    \hfill
    \begin{subfigure}
        \centering
        \includegraphics[width=3.3in]{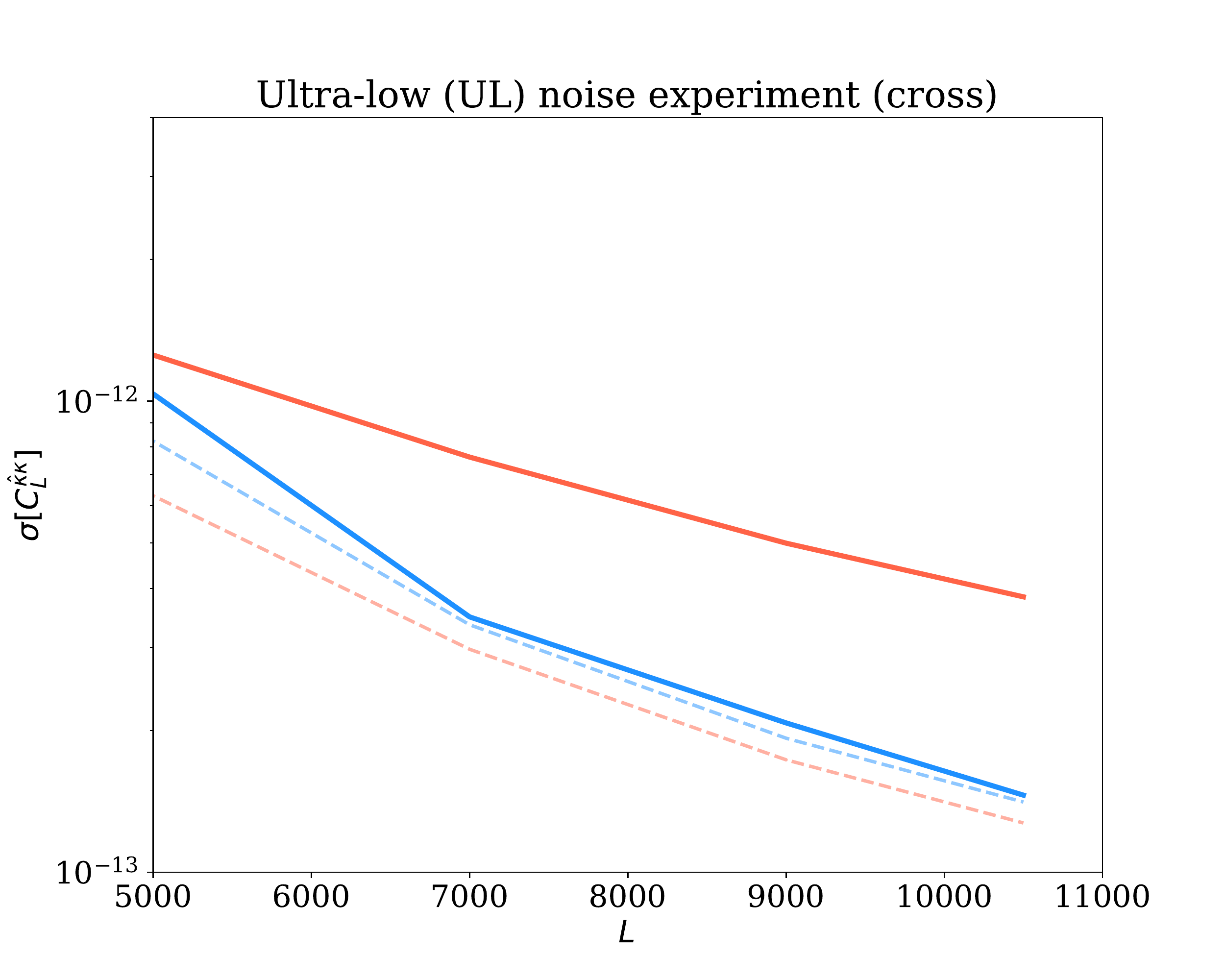}
    \end{subfigure}
    \caption{Standard deviation of the
    auto (\textit{upper panels}) and cross 
      (\textit{lower panels}) power spectra
      from multiple simulations for an
     experiment with CMB-S4-like 
     (\textit{left plots}) and ultra-low
     noise (\textit{right plots}). QE results are in red, while GI results are in blue.  The thick lines denote simulation, while the fainter dashed ones are derived from a theoretical prediction of the noise.
    Here we discard the $L < 5000$ modes, where our method is expected to be suboptimal. The error bars are
    scaled such that they simulate a sky coverage of 40\%. The QE theory curves are ``naive'' predictions that ignore mode correlations, while the GI theory, which shows good agreement with the simulations results,
    curves take into account the
    effective area left unmasked
    by the inverse-noise filter. In (\textit{dotted black}) we denote the power spectrum of the convergence field (signal). The bin width here is $\Delta L = 2000$.}
\label{fig:astd}
\end{figure*}

For this purpose, we will use simulations to compute the errors on measurements of auto- and cross-power spectra with the reconstructed lensing field. Using the 360 simulations of lensing reconstruction described in the previous section, we measure and bin the lensing power spectrum and the lensing cross-correlation with the input convergence field; from these simulated measurements, we determine the covariance matrices and standard deviations for cross- and auto-spectra, for both GI and QE. For the autospectra, we assume that reconstruction noise biases will be characterized by monte-carlo simulations; since subtracting a mean bias does not affect fluctuations, we do not subtract the reconstruction noise bias when determining covariances. We show our results in Fig. \ref{fig:astd}.

As seen in the top panels of Fig. \ref{fig:astd}, the standard
deviation (error) of the measured \textbf{auto}-power 
spectrum for the ultra-low noise experiment 
(top right) using
the GI estimator is significantly
smaller than the result from the QE
simulations across all $L$-modes under consideration,
reaching a full order of magnitude in difference around $L \sim 10000$. 
For an experiment with CMB-S4-like noise (top left),
the error in the GI auto-correlation measurements
is smaller by about a factor of 5 at $L \sim 6500$
compared with the QE, and seems to decrease the smaller the scale
probed, leveling off and remaining non-negligible.

The bottom two panels of Fig. \ref{fig:astd} 
shows the standard deviation of the measured
\textbf{cross}-power 
spectrum. In both the case of an ultra-low noise experiment
and a CMB-S4-like experiment, the GI estimator appears to have
smaller error bars
than the QE on small scales ($L \gtrsim 4000$).
In particular, for the UL-noise case 
(light-red and light-green curves),
on scales around $L \gtrsim 10000$, our
method outperforms the QE significantly, improving it by a factor of
$\sim 4$. The CMB-S4 case 
(bottom left panel) also sees improvement when
the GI method is applied, but it is somewhat more modest, at about
a factor of $\sim 2$ over all scales shown.

We can also compare our simulated results with approximate theoretical predications. In Section \ref{sec:theo}, we presented the theoretical 
framework for forcasting the noise of the GI approach. In particular, we showed that the noise on our estimator
(in terms of the lensing potential field)
approximately takes the form given in Eq. \ref{eq:noise_mode}
(see a more detailed discussion of the GI noise in Appendix \ref{app:noise}).
We   assume Gaussian covariance \cite{knox} to convert this reconstruction noise power to errors $\sigma^2[C_L^{AB}]$:
\begin{equation}
  \sigma^2[C_L^{AB}] = \frac{1}{\Delta L (2L+1) f_{\rm sky}}(C_L^{AA}C_L^{BB}+[C_L^{AB}]^2),
  \label{eq:knoxFormula}
\end{equation}
where the power spectra $C_L^{AB}$ can include cross-spectra $C_L^{\hat \kappa \kappa}$ and autospectra $C_L^{\hat \kappa \hat \kappa}$, which include the reconstruction noise power. Following our results in Fig.~\ref{fig:cl}, we assume $\langle C_L^{\hat \kappa \kappa}\rangle = C_L^{ \kappa \kappa}$. We thus obtain theoretical predictions for the errors which are shown in Fig.~\ref{fig:astd} using thin lines.

We expect our theoretically derived expression for
the GI noise to be accurate
in the limit $L > 4000$ from Fig. \ref{fig:cl},
where the primary CMB is
highly suppressed by diffusion damping. This is
indeed what is observed in Fig. \ref{fig:astd}:
for ($L \gtrsim 6500$), GI theory and simulation
are in good agreement with each other for the case of
CMB-S4-like experiment (dark-blue thin vs.~thick curves).
For the UL-noise experiment the agreement becomes more evident
at $L \gtrsim 9000$.
On larger scales, $L \lesssim 4000$, the primordial
temperature fluctuations are not negligible, and many assumptions made in deriving the GI estimator on small scales break down. Therefore, it is perhaps not surprising that the errors from
the simulations do not match the approximate theoretical calculation on these scales.
Regarding the match of simulated results to simple theoretical predictions for the QE,
as shown in Nguyen et al. \cite{Nguyen:2017zqu} for bandpowers (and in Horowitz et al. \cite{Horowitz:2017iql} for clusters)  ,
we confirm that the QE indeed has a much larger error on small scales
than expected
by the ``naive'' prediction which neglects mode coupling.  
This is especially true of the 
UL-noise experiment, where the difference is of about an order of
magnitude on all scales. 
This suggests that if one wants to forecast the QE noise more correctly, a more careful treatment of its reconstruction noise has to be applied.  On larger scales ($L <  1000$), we have verified that QE theory and simulation agree well, as expected. We emphasize that the large improvements of the GI estimator we have found are relative to the simulated QE, not the QE forecasts (which are overly optimistic).

\begin{table}[ht]
\centering
\begin{tabular}{||c || c| c| c||} 
 \hline
 SNR & UL & S4-like & SO-like   \\ [0.5ex] 
 \hline\hline
 $L$-range & 4000 $-$ 18000 & 5000 $-$ 11000  & 5000 $-$ 9000 \\
 \hline
 Cross QE &  710 & 550 & 195 \\ 
 Cross GI & 4100 & 1440 & 270 \\
 \hline
 Auto QE &  205 & 100 & 7 \\ 
 Auto GI & 1515 & 360 & 30 \\ [1ex]
 \hline
\end{tabular}
\caption{Signal-to-noise ratio (SNR) computed from simulations ($\sim$300 for GI and 1000 for QE) of experiments with three different noise-levels: ultra-low noise (UL), CMB-Stage-IV-like (S4) and Simons-Observatory-like (SO).}
\label{table:1}
\label{tab:snr}
\end{table}

To summarize the improvements expected from the GI estimator, we now calculate the cumulative signal-to-noise on lensing observables for different experiments. In Table \ref{tab:snr}, we show
the signal-to-noise ratio, 
\be
{\rm SNR}^2  = \sum_{ij} C_L^{i \ \kappa \kappa}\widehat{\rm  Cov}^{-1}_{ij} C_L^{j \ \kappa \kappa},
\ee
computed over a range of 
$L$-modes for three noise 
levels, mimicking the fiducial
ultra-low noise and 
Stage-IV-like experiments, with $i$ and $j$ indexing the bandpower bins and $\widehat{\rm  Cov}$
denoting the covariance of the 
measured power spectra, $C_L^{ \hat \kappa \kappa}$ (cross) and
$C_L^{\hat \kappa \hat \kappa}$ (auto). Notice that here we loosen
the assumption of Gaussian 
covariance to obtain more accurate estimates of the SNR, as indeed
we observe non-negligible
correlations between the 
small-scale modes of the QE-reconstructed power spectra. This is less true in the GI case, as expected.
We have also added a forecast
for an experiment similar to Simons Observatory (SO)  , assuming instrumental 
noise with beam $\theta_{\rm FWHM} = 1.4$ arcmin and noise factor
$\Delta_{\rm T}=5\mu$K-arcmin, as it is the fastest approaching 
 CMB observational
project with potential to go to small enough scales in temperature. This calculation also includes the effect of off-diagonal bandpower covariances. We infer an
improvement of the GI
estimator over the QE on the order of
$\sim 5-6$ times for the UL-noise case
and about $\sim 2.5$ and $\sim 3$ for SO-like and CMB-S4,
respectively, on small scales $L=5000-9000$.

We note that applying a more optimal window function, such as the full Wiener filter in Eq. \ref{mainEstimator},
 could, to some extent, further increase the precision of the GI lensing measurements on small-scales. 

In the small-scale regime, we caution that there are significant contributions from foreground
contaminants which still need to be properly accounted for (see Subsection \ref{subsec:sys}).
Though there are several possibilities for mitigating these with multifrequency cleaning and estimator modification, we note that our forecasts are therefore likely optimistic. Nevertheless, especially at moderate $L$, the GI estimator appears promising for improving lensing signal-to-noise from future experiments beyond what was thought possible using the QE estimator.

\subsection{Effective Reconstruction Noise}
Another useful metric for comparing estimator performance is to calculate the reconstruction noise power spectrum, or, equivalently, the noise per lensing mode. However, due to inhomogeneous spatial weighting, the GI estimator can have a different effective area than the QE, so that a naive comparison is difficult. We proceed instead as follows. We define the effective reconstruction noise $ N_{L, {\rm eff}}^{\kappa \kappa}$ as the level of noise that would give the same power spectrum error bars given a simple forecast with uniform weighting across the map. We can therefore calculate it by inverting the Gaussian covariance   formula defined above (Eq.~\ref{eq:knoxFormula}) for the lensing autospectrum to yield
\begin{equation}
  N_{L, {\rm eff}}^{\kappa \kappa} \equiv \hat \sigma[C_L^{\hat \kappa \hat \kappa}] \sqrt{\Delta L (L+\frac{1}{2}) f_{\rm sky}}-C_L^{\kappa \kappa},
    \label{eq:eff_noise}
\end{equation}
where $\hat \sigma[C_L^{\hat \kappa \hat \kappa}]$ is the standard deviation of the auto power spectra measured from multiple
simulations and $C_L^{\kappa \kappa}$ is the theoretical power
spectrum of the convergence field.

\begin{figure}
  \centering  \includegraphics[width=0.5\textwidth]{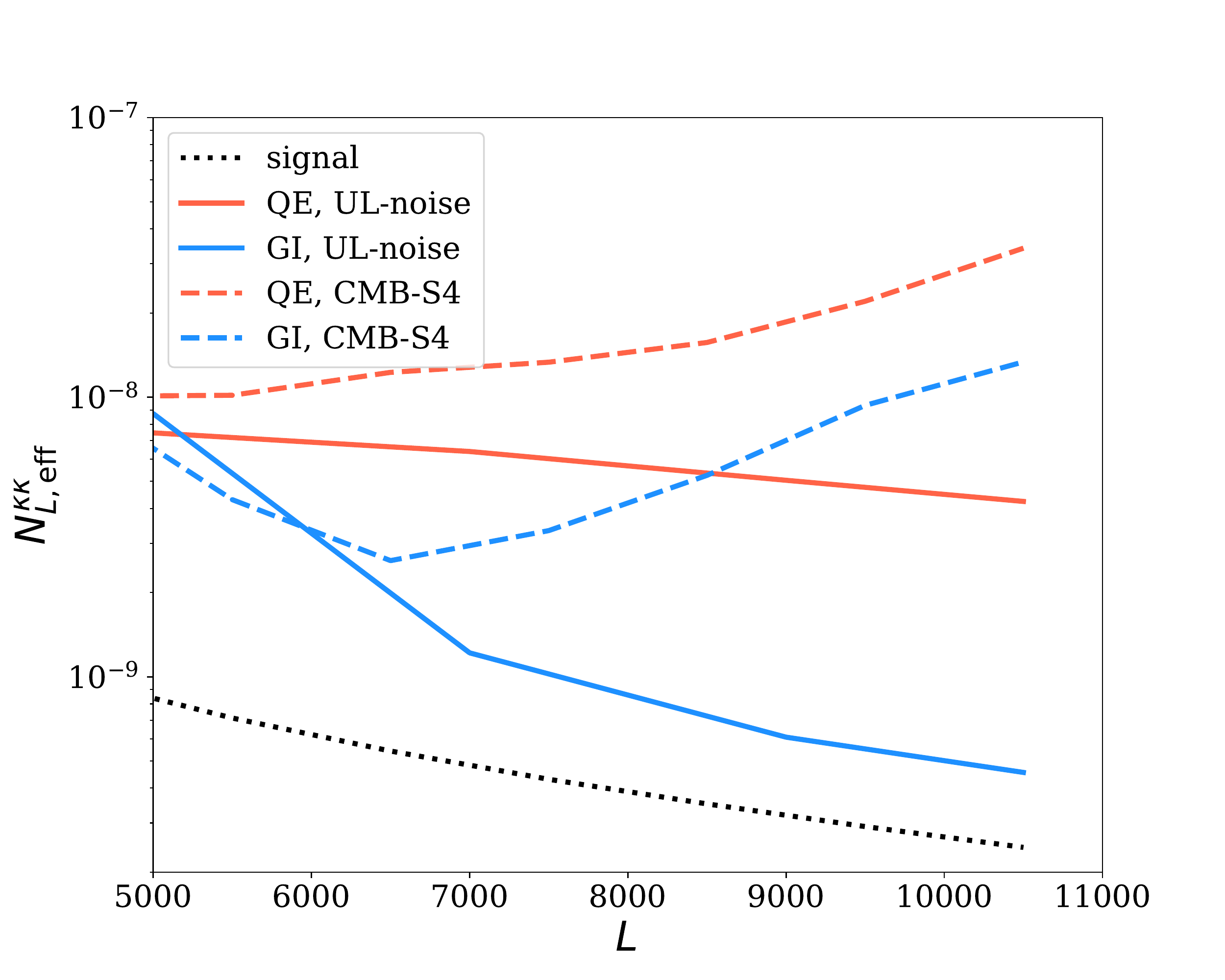}
    \caption{Effective noise curves for an ultra-low noise and CMB-S4-like
      experiment from simulations.
      The largest improvement occurs for wave modes around 
$L \sim 10000$ of the order of $\sim$ 10 for the UL-noise case and $\sim 5$ around $L \sim 6500$ for the CMB-S4 case. The effective noise is inferred from the scatter in the bandpowers of the auto-spectrum of each each estimator as in Eq. \ref{eq:eff_noise}. }
    \label{fig:nl}
\end{figure}

In Fig. \ref{fig:nl},
 we show the QE and GI effective noise curves,
where we use statistical results derived from simulations (360
GI reconstructions for UL noise; 300 GI reconstructions for SO-like and S4-like noise; 1000 QE reconstructions for each noise level).
The graphs
suggest that the GI noise of the estimator for
an ultra-low noise experiment is on the order
of 10 times smaller than that of the QE for
$L \sim 10000$. For an experiment with
CMB-S4-like instrumental noise, the improvement is more modest,
peaking at $L \approx 6500$, where the
ratio between the two curves is about 5. In both cases, the GI estimator results in a significantly more accurate measurement of the lensing field.

\subsection{Systematic Errors from Foregrounds}
\label{subsec:sys}
Though the methods we have described for improved small-scale lensing reconstruction appear promising, small-scale extragalactic foregrounds will likely be a significant limiting factor for such analyses.

This is especially true for temperature reconstruction, which is the focus of this paper and is expected to have higher signal-to-noise. Polarization reconstruction should have minimal small-scale foregrounds, and we will briefly discuss the extension of our methods to polarization in Appendix \ref{app:polarization}, but we will defer a detailed treatment to future work.

In temperature, as previously mentioned, several extragalactic foregrounds are of primary concern on small angular scales: Cosmic Infrared Background (CIB) emission and the thermal and kinematic Sunyaev-Zel'dovich effects (tSZ and kSZ). Other contaminants such as radio source emission are likely subdominant. CIB and tSZ foregrounds have a distinctive frequency dependence; this, in principle, allows them to be separated from the blackbody CMB and lensing signals using multi-frequency
measurements of the small-scale CMB from upcoming experiments. 

The kSZ effect, on the other hand, cannot be separated using multifrequency data, as it has a blackbody frequency dependence. At first glance it therefore presents a more serious challenge, with the potential to significantly bias small-scale lensing measurements. However, the kSZ differs in an important way from lensing: the lensing signal is correlated with the background CMB gradient, whereas the kinetic SZ effect and other foregrounds are not. (This, of course, assumes that the gradient itself has a negligible kSZ contribution; we will briefly revisit this assumption below.) Building on this statistical difference, we can construct simple methods to avoid any kSZ bias to lensing measurements.

We will here briefly outline such a method to remove kSZ bias in autospectrum measurements using the GI lensing estimator. To calculate the bias from kSZ (in addition to that arising from instrument noise and other sources), we simply evaluate the GI estimator with the gradient coordinates offset by a small vector $\br$,
\begin{eqnarray}
&&{\hat \phi^\bL}_{bias} (\bx)  \approx\\ &&W^\bL( \bx+\br)\frac{ \tl(\bx)_{\text{high-pass}}}{(i\bL \cdot \nabla T(\bx+\br))} \times \left[\left < W^\bL( \bx+\br) \right >_\bx \right]^{-1} . \nonumber
\end{eqnarray}
Here $\br$ is chosen to be longer than the correlation lengths of the CMB fluctuations and the lensing, but small enough that the noise properties are unchanged. Since the kSZ, noise and
other effects and foregrounds are not correlated with the large scale gradient \cite{Ferraro:2017fac,Smith:2016lnt,Amblard:2004ih}, we can determine the non-lensing foreground and noise bias by simply calculating the power spectrum of $
{\hat \phi^\bL}_{bias}$. This can be subtracted off from the raw reconstruction power to calculate the lensing power spectrum, taking $(C^{\hat \phi \hat \phi}_\bL-C^{\hat \phi \hat \phi}_{\bL, bias})\times N_\mathbf{L}$. The normalization
function $N_\mathbf{L}$ (which can be obtained via simulation or analytics) is required because $C^{\hat \phi \hat \phi}_{\bL, bias}$ also contains some lensing signal, though it is very sub-optimally weighted and thus significantly reduced in amplitude; re-normalization is needed to restore an unbiased measurement of the lensing power spectrum.

While this method should remove the leading source of kSZ bias, an additional, smaller non-Gaussian bias could arise because the measured CMB gradient also contains kSZ at a low level; this should be explored further in future work.

We also note that, even with reduced foreground biases, the kSZ contribution could still inflate errors and lower signal-to-noise in temperature, especially at very high $L$. Possible ways to reduce kSZ foreground contributions to errors (aside from simply relying on polarization-based reconstructions) might include subtracting a reconstructed kSZ map derived from galaxy surveys or quadratic kSZ reconstructions. We will defer a detailed exploration of foreground issues to future work.

\section{Conclusions}
The standard tool used for reconstruction of the gravitational lensing field
is the quadratic estimator, but it has recently been shown to be suboptimal on small angular scales. 
The most rigorous way to overcome this limitation is to adopt an optimal approach (e.g., \cite{hirata2003analyzing,carron2017maximum, Millea:2017fyd}), but this is not necessarily the most computationally efficient one, and in any case, algorithms that rapidly converge on small scales have not yet been presented.
In this small-scale regime, a simple gradient-inversion estimator (GI) should approach the optimal
solution. 

In this paper, we have derived a more general GI algorithm, one that is now capable of reconstructing lensing over large fields. We have applied it to simulations of the lensed CMB temperature field to validate the method and characterize its performance.
For idealized, foreground-free simulations, we have shown that
the GI-reconstructed auto and cross correlation error bars are significantly smaller than those
derived using the QE. In particular, we found that when using our method  on small scales for the computation
and analysis of the cross-correlations 
between these next-generation experiments, one obtains a much more precise result,
with measurement errors which are approximately $\sim 2.6$ times smaller
for CMB-S4 and $\sim 1.4$ for SO-like than the corresponding
ones for QE at $L=5000-9000$. The improvement of the auto-correlation
power spectrum is even more substantial: a factor of $\sim 4$ for both 
CMB-S4 and SO-like in the highly optimistic scenario of removed foregrounds and measured small-scale anisotropies. We argue that the origin of these improvements lies in the fact that, unlike the GI,
the QE is limited by gradient cosmic variance. 

We expect that the small-scale lensing measurements could, to some extent, be even further improved by fully implementing the weighting derived in Eq.\ref{mainEstimator} without any approximations.

Though we have presented simple ideas for mitigating or removing foreground biases to the estimator, we caution that the levels of extragalactic foregrounds are large, and that a full examination of the impact of foregrounds has been left to future work. However, as sketched out in Appendix \ref{app:polarization}, an analogous version of the GI estimator can also easily
 be derived for the case of polarization; the polarization GI
estimator has the advantage that it does not suffer from comparable foreground contamination.

Small-scale lensing reconstruction has many potential applications
to cosmology such as constraining
cosmological parameters, distinguishing between dark matter models (e.g. \cite{Nguyen:2017zqu},\cite{CMBHD}), validating galaxy weak lensing shear measurements, and constraining high-redshift astrophysics. If further work on polarization, foregrounds and systematics is successful, the improved small-scale lensing reconstruction algorithms presented here could allow new scientific possibilities in these areas.

\acknowledgments
We thank Anthony Challinor, Ben Horowitz, Antony Lewis, Emmanuel Schaan and Uros Seljak for useful discussions. 
B.H. was supported by the Benefactors'
scholarship at St John's College, Cambridge University.
B.D.S. was supported by an STFC Ernest Rutherford Fellowship and an Isaac Newton Trust Early Career Grant.
S.F. was in part supported by a Miller Fellowship at the University of California, Berkeley and by the Physics Division at Lawrence Berkeley National Laboratory.

\appendix

\section{Noise Discussion}
\label{app:noise}

In this appendix, we discuss the theoretical
prediction of the reconstruction noise in more detail.
As discussed in Section \ref{sec:theo}, the noise on our estimator for a local gradient is given by
\begin{equation}
N_\bL^{\hat \pot \hat \pot}(\bx) = \frac{N_L^{TT}}{(\bL \cdot \nabla T(\bx))^2}.
\end{equation}

We average over space and over different 
gradients to obtain the noise power spectrum.
If the gradient modes are uncorrelated,
then the minimum variance weights are given by
\begin{equation}
N_L^{\hat \pot \hat \pot} 
 = \left[ \int \frac{d^2 \bx}{A} \ \frac{1}{N_\bL^{\hat \pot \hat \pot}( \bx)} \right]^{-1} = \frac{2 N_L^{TT}}{L^2 |\nabla T|^2_{\rm rms}},
\label{eq:noise}
\end{equation}
where $A$ is the total area over which we are averaging.

In our initial approximations we have assumed that the noise on the CMB temperature on small scales is merely the instrumental noise; in reality, it can be more accurately modeled by ``noise'' power coming from other first-order contributions to the lensed temperature power 
spectrum (see Eq. 4.16 in 
Ref.~\cite{Lewis:2006fu})
\be
N_L^{TT} = C_L^{\rm noise} + C_L^{\tl \tl}- \frac{1}{2} L^2 C_L^{\pot \pot} |\nabla T|_{\rm rms}^2.
\ee
The second and third terms here are an attempt to isolate the contributions to the lensed power spectrum that do not arise from a particular lensing mode.
The instrumental noise is estimated using the conventional parametrization
\cite{knox} to be
\be
C_L^{\rm noise}
= \Delta_{\rm T}
^2 \exp\left[\frac{L (L+1)\theta^2_{\rm FWHM}}{8 \ln 2}\right],
\ee
where the noise factor, $\Delta_{\rm T}$, is measured in
$\mu$K-rad and the beam full-width half
maximum (FWHM),
$\theta_{\rm FWHM}$,
in radians 
(given in arcmin in the paper for convenience).

\section{Polarization Estimator}
\label{app:polarization}
Similarly to the temperature estimators 
considered earlier, we can straightforwardly 
extend the GI method to polarization. In this
appendix, we outline the approach one would take,
assuming a constant gradient within the patch
under consideration. This approximation is in 
fact not completely unreasonable
in the limit of small-scale lensing
we are interested in. A proper treatment of 
combining different patches with different
gradient magnitudes and directions would follow
in an analogous manner to what was developed in
Section \ref{sec:theo}.

As shown in \cite{Lewis:2006fu}, we can write
the lensed $E$ and $B$ fields on small scales as:
\beq
\tilde E(\bL) \approx  n(\bL)+E(\bL)- \phi(\bL) \bL \cdot(\nabla Q \cos(2 \varphi_\bL)+\nabla U \sin(2 \varphi_\bL)) 
\label{eq:te}
\eeq
\beq
\tilde B(\bL) \approx  n(\bL)+B(\bL)- \phi(\bL) \bL \cdot(-\nabla Q \sin(2 \varphi_\bL)+\nabla U \cos(2 \varphi_\bL)) .
\label{eq:tb}
\eeq
We then have two estimators
for the deflection potential with two filter functions:
\beq
\hat \phi_1 (\bL) \equiv  \tilde E(\bL) F_1 (\bL)
\eeq
\beq
\hat \phi_2 (\bL) \equiv  \tilde B(\bL) F_2 (\bL) .
\eeq
Multiplying Eq. \ref{eq:te} and Eq. \ref{eq:tb} 
by $F_1 (\bL)$ and $F_2 (\bL)$, 
we average over CMB realizations of the polarization map, requiring that the estimators are unbiased,
to obtain:
\begin{align*}
    \phi (\bL) &= \langle \hat \phi_1(\bL) \rangle_{\rm CMB} = \langle \tilde E(\bL) F_1(\bL) \rangle_{\rm CMB} = \nonumber \\ 
    &= 
\langle \phi(\bL) \bL \cdot(\nabla Q \cos(2 \varphi_\bL)+\nabla U \sin(2 \varphi_\bL)) F_1(\bL) \rangle_{\rm CMB} 
\end{align*}
\begin{align*}
    \phi(\bL) &= \langle \hat \phi_2(\bL) \rangle_{\rm CMB} = \langle \tilde B(\bL) F_2(\bL) \rangle_{\rm CMB} = \nonumber \\ 
    &=
\langle \phi(\bL) \bL \cdot(-\nabla Q \sin(2 \varphi_\bL)+\nabla U \cos(2 \varphi_\bL)) F_2(\bL) \rangle_{\rm CMB} .
\end{align*}
It follows that:
\beq
F_1(\bL) = \frac{1}{\bL \cdot(\nabla Q \cos(2 \varphi_\bL)+\nabla U \sin(2 \varphi_\bL))} 
\eeq
\beq
F_2(\bL) = \frac{1}{\bL \cdot(-\nabla Q \sin(2 \varphi_\bL)+\nabla U \cos(2 \varphi_\bL))} .
\eeq

We, thus, end up with the two estimators:
\begin{align*}
    \hat \phi_1 (\bL)  &= \tilde E(\bL) F_1(\bL) 
    \nonumber \\
    &= \frac{n(\bL)+E(\bL)+\bL \cdot(\nabla Q \cos(2 \varphi_\bL)+\nabla U \sin(2 \varphi_\bL))}{\bL \cdot(\nabla Q \cos(2 \varphi_\bL)+\nabla U \sin(2 \varphi_\bL))}
\end{align*}
\begin{align*}
\hat \phi_2(\bL)  &= \tilde B(\bL) F_2(\bL)
\nonumber \\
&= \frac{n(\bL)+\bL \cdot(-\nabla Q \sin(2 \varphi_\bL)+\nabla U \cos(2 \varphi_\bL))}{\bL \cdot(-\nabla Q \sin(2 \varphi_\bL)+\nabla U \cos(2 \varphi_\bL))},
\end{align*}
with noise power spectra:
\beq
N^{1,\hat \phi \hat \phi}_L=\langle (\hat \phi_1 - \phi)^2 \rangle=\frac{N_L^{EE}}{ [\bL \cdot(\nabla Q \cos(2 \varphi_\bL)+\nabla U \sin(2 \varphi_\bL))]^2 }
\eeq
\\
\beq
N^{2,\hat \phi \hat \phi}_L=\langle (\hat \phi_2 - \phi)^2 \rangle=\frac{N_L^{BB}}{ [\bL \cdot(-\nabla Q \sin(2 \varphi_\bL)+\nabla U \cos(2 \varphi_\bL))]^2 }.
\eeq

We thus effectively get two new estimators for 
the lensing potential which weighted appropriately, can provide a better estimator
for the lensing potential. Polarization
measurements, in
addition, have significantly lower foreground contamination compared to temperature 
measurements. The generalization for many
CMB patches of different gradients follows
the approach outlined in Section \ref{sec:theo}
and will be addressed in a subsequent paper.

\bibliographystyle{apsrev4-1}     
\bibliography{main} 

\begin{thebibliography}{28}%
\makeatletter
\providecommand \@ifxundefined [1]{%
 \@ifx{#1\undefined}
}%
\providecommand \@ifnum [1]{%
 \ifnum #1\expandafter \@firstoftwo
 \else \expandafter \@secondoftwo
 \fi
}%
\providecommand \@ifx [1]{%
 \ifx #1\expandafter \@firstoftwo
 \else \expandafter \@secondoftwo
 \fi
}%
\providecommand \natexlab [1]{#1}%
\providecommand \enquote  [1]{``#1''}%
\providecommand \bibnamefont  [1]{#1}%
\providecommand \bibfnamefont [1]{#1}%
\providecommand \citenamefont [1]{#1}%
\providecommand \href@noop [0]{\@secondoftwo}%
\providecommand \href [0]{\begingroup \@sanitize@url \@href}%
\providecommand \@href[1]{\@@startlink{#1}\@@href}%
\providecommand \@@href[1]{\endgroup#1\@@endlink}%
\providecommand \@sanitize@url [0]{\catcode `\\12\catcode `\$12\catcode
  `\&12\catcode `\#12\catcode `\^12\catcode `\_12\catcode `\%12\relax}%
\providecommand \@@startlink[1]{}%
\providecommand \@@endlink[0]{}%
\providecommand \url  [0]{\begingroup\@sanitize@url \@url }%
\providecommand \@url [1]{\endgroup\@href {#1}{\urlprefix }}%
\providecommand \urlprefix  [0]{URL }%
\providecommand \Eprint [0]{\href }%
\providecommand \doibase [0]{http://dx.doi.org/}%
\providecommand \selectlanguage [0]{\@gobble}%
\providecommand \bibinfo  [0]{\@secondoftwo}%
\providecommand \bibfield  [0]{\@secondoftwo}%
\providecommand \translation [1]{[#1]}%
\providecommand \BibitemOpen [0]{}%
\providecommand \bibitemStop [0]{}%
\providecommand \bibitemNoStop [0]{.\EOS\space}%
\providecommand \EOS [0]{\spacefactor3000\relax}%
\providecommand \BibitemShut  [1]{\csname bibitem#1\endcsname}%
\let\auto@bib@innerbib\@empty
\bibitem [{\citenamefont {{Smith}}\ \emph {et~al.}(2006)\citenamefont
  {{Smith}}, \citenamefont {{Hu}},\ and\ \citenamefont
  {{Kaplinghat}}}]{2006PhRvD..74l3002S}%
  \BibitemOpen
  \bibfield  {author} {\bibinfo {author} {\bibfnamefont {K.~M.}\ \bibnamefont
  {{Smith}}}, \bibinfo {author} {\bibfnamefont {W.}~\bibnamefont {{Hu}}}, \
  and\ \bibinfo {author} {\bibfnamefont {M.}~\bibnamefont {{Kaplinghat}}},\
  }\href {\doibase 10.1103/PhysRevD.74.123002} {\bibfield  {journal} {\bibinfo
  {journal} {\prd}\ }\textbf {\bibinfo {volume} {74}},\ \bibinfo {eid} {123002}
  (\bibinfo {year} {2006})},\ \Eprint {http://arxiv.org/abs/astro-ph/0607315}
  {arXiv:astro-ph/0607315 [astro-ph]} \BibitemShut {NoStop}%
\bibitem [{\citenamefont {{Planck Collaboration}}\ \emph
  {et~al.}(2018)\citenamefont {{Planck Collaboration}}, \citenamefont
  {{Aghanim}}, \citenamefont {{Akrami}}, \citenamefont {{Ashdown}},
  \citenamefont {{Aumont}}, \citenamefont {{Baccigalupi}}, \citenamefont
  {{Ballardini}}, \citenamefont {{Banday}}, \citenamefont {{Barreiro}},
  \citenamefont {{Bartolo}}, \citenamefont {{Basak}}, \citenamefont
  {{Benabed}}, \citenamefont {{Bernard}}, \citenamefont {{Bersanelli}},
  \citenamefont {{Bielewicz}}, \citenamefont {{Bock}}, \citenamefont {{Bond}},
  \citenamefont {{Borrill}}, \citenamefont {{Bouchet}}, \citenamefont
  {{Boulanger}}, \citenamefont {{Bucher}}, \citenamefont {{Burigana}},
  \citenamefont {{Calabrese}}, \citenamefont {{Cardoso}}, \citenamefont
  {{Carron}}, \citenamefont {{Challinor}}, \citenamefont {{Chiang}},
  \citenamefont {{Colombo}}, \citenamefont {{Combet}}, \citenamefont {{Crill}},
  \citenamefont {{Cuttaia}}, \citenamefont {{de Bernardis}}, \citenamefont {{de
  Zotti}}, \citenamefont {{Delabrouille}}, \citenamefont {{Di Valentino}},
  \citenamefont {{Diego}}, \citenamefont {{Dor{\'e}}}, \citenamefont
  {{Douspis}}, \citenamefont {{Ducout}}, \citenamefont {{Dupac}}, \citenamefont
  {{Efstathiou}}, \citenamefont {{Elsner}}, \citenamefont {{En{\ss}lin}},
  \citenamefont {{Eriksen}}, \citenamefont {{Fantaye}}, \citenamefont
  {{Fernandez-Cobos}}, \citenamefont {{Forastieri}}, \citenamefont {{Frailis}},
  \citenamefont {{Fraisse}}, \citenamefont {{Franceschi}}, \citenamefont
  {{Frolov}}, \citenamefont {{Galeotta}}, \citenamefont {{Galli}},
  \citenamefont {{Ganga}}, \citenamefont {{G{\'e}nova-Santos}}, \citenamefont
  {{Gerbino}}, \citenamefont {{Ghosh}}, \citenamefont {{Gonz{\'a}lez-Nuevo}},
  \citenamefont {{G{\'o}rski}}, \citenamefont {{Gratton}}, \citenamefont
  {{Gruppuso}}, \citenamefont {{Gudmundsson}}, \citenamefont {{Hamann}},
  \citenamefont {{Hand ley}}, \citenamefont {{Hansen}}, \citenamefont
  {{Herranz}}, \citenamefont {{Hivon}}, \citenamefont {{Huang}}, \citenamefont
  {{Jaffe}}, \citenamefont {{Jones}}, \citenamefont {{Karakci}}, \citenamefont
  {{Keih{\"a}nen}}, \citenamefont {{Keskitalo}}, \citenamefont {{Kiiveri}},
  \citenamefont {{Kim}}, \citenamefont {{Knox}}, \citenamefont
  {{Krachmalnicoff}}, \citenamefont {{Kunz}}, \citenamefont {{Kurki-Suonio}},
  \citenamefont {{Lagache}}, \citenamefont {{Lamarre}}, \citenamefont
  {{Lasenby}}, \citenamefont {{Lattanzi}}, \citenamefont {{Lawrence}},
  \citenamefont {{Le Jeune}}, \citenamefont {{Levrier}}, \citenamefont
  {{Lewis}}, \citenamefont {{Liguori}}, \citenamefont {{Lilje}}, \citenamefont
  {{Lindholm}}, \citenamefont {{L{\'o}pez-Caniego}}, \citenamefont {{Lubin}},
  \citenamefont {{Ma}}, \citenamefont {{Mac{\'\i}as-P{\'e}rez}}, \citenamefont
  {{Maggio}}, \citenamefont {{Maino}}, \citenamefont {{Mandolesi}},
  \citenamefont {{Mangilli}}, \citenamefont {{Marcos-Caballero}}, \citenamefont
  {{Maris}}, \citenamefont {{Martin}}, \citenamefont
  {{Mart{\'\i}nez-Gonz{\'a}lez}}, \citenamefont {{Matarrese}}, \citenamefont
  {{Mauri}}, \citenamefont {{McEwen}}, \citenamefont {{Melchiorri}},
  \citenamefont {{Mennella}}, \citenamefont {{Migliaccio}}, \citenamefont
  {{Miville-Desch{\^e}nes}}, \citenamefont {{Molinari}}, \citenamefont
  {{Moneti}}, \citenamefont {{Montier}}, \citenamefont {{Morgante}},
  \citenamefont {{Moss}}, \citenamefont {{Natoli}}, \citenamefont {{Pagano}},
  \citenamefont {{Paoletti}}, \citenamefont {{Partridge}}, \citenamefont
  {{Patanchon}}, \citenamefont {{Perrotta}}, \citenamefont {{Pettorino}},
  \citenamefont {{Piacentini}}, \citenamefont {{Polastri}}, \citenamefont
  {{Polenta}}, \citenamefont {{Puget}}, \citenamefont {{Rachen}}, \citenamefont
  {{Reinecke}}, \citenamefont {{Remazeilles}}, \citenamefont {{Renzi}},
  \citenamefont {{Rocha}}, \citenamefont {{Rosset}}, \citenamefont {{Roudier}},
  \citenamefont {{Rubi{\~n}o-Mart{\'\i}n}}, \citenamefont {{Ruiz-Granados}},
  \citenamefont {{Salvati}}, \citenamefont {{Sandri}}, \citenamefont
  {{Savelainen}}, \citenamefont {{Scott}}, \citenamefont {{Sirignano}},
  \citenamefont {{Sunyaev}}, \citenamefont {{Suur-Uski}}, \citenamefont
  {{Tauber}}, \citenamefont {{Tavagnacco}}, \citenamefont {{Tenti}},
  \citenamefont {{Toffolatti}}, \citenamefont {{Tomasi}}, \citenamefont
  {{Trombetti}}, \citenamefont {{Valiviita}}, \citenamefont {{Van Tent}},
  \citenamefont {{Vielva}}, \citenamefont {{Villa}}, \citenamefont
  {{Vittorio}}, \citenamefont {{Wandelt}}, \citenamefont {{Wehus}},
  \citenamefont {{White}}, \citenamefont {{White}}, \citenamefont {{Zacchei}},\
  and\ \citenamefont {{Zonca}}}]{2018arXiv180706210P}%
  \BibitemOpen
  \bibfield  {author} {\bibinfo {author} {\bibnamefont {{Planck
  Collaboration}}}, \bibinfo {author} {\bibfnamefont {N.}~\bibnamefont
  {{Aghanim}}}, \bibinfo {author} {\bibfnamefont {Y.}~\bibnamefont {{Akrami}}},
  \bibinfo {author} {\bibfnamefont {M.}~\bibnamefont {{Ashdown}}}, \bibinfo
  {author} {\bibfnamefont {J.}~\bibnamefont {{Aumont}}}, \bibinfo {author}
  {\bibfnamefont {C.}~\bibnamefont {{Baccigalupi}}}, \bibinfo {author}
  {\bibfnamefont {M.}~\bibnamefont {{Ballardini}}}, \bibinfo {author}
  {\bibfnamefont {A.~J.}\ \bibnamefont {{Banday}}}, \bibinfo {author}
  {\bibfnamefont {R.~B.}\ \bibnamefont {{Barreiro}}}, \bibinfo {author}
  {\bibfnamefont {N.}~\bibnamefont {{Bartolo}}}, \bibinfo {author}
  {\bibfnamefont {S.}~\bibnamefont {{Basak}}}, \bibinfo {author} {\bibfnamefont
  {K.}~\bibnamefont {{Benabed}}}, \bibinfo {author} {\bibfnamefont {J.~P.}\
  \bibnamefont {{Bernard}}}, \bibinfo {author} {\bibfnamefont {M.}~\bibnamefont
  {{Bersanelli}}}, \bibinfo {author} {\bibfnamefont {P.}~\bibnamefont
  {{Bielewicz}}}, \bibinfo {author} {\bibfnamefont {J.~J.}\ \bibnamefont
  {{Bock}}}, \bibinfo {author} {\bibfnamefont {J.~R.}\ \bibnamefont {{Bond}}},
  \bibinfo {author} {\bibfnamefont {J.}~\bibnamefont {{Borrill}}}, \bibinfo
  {author} {\bibfnamefont {F.~R.}\ \bibnamefont {{Bouchet}}}, \bibinfo {author}
  {\bibfnamefont {F.}~\bibnamefont {{Boulanger}}}, \bibinfo {author}
  {\bibfnamefont {M.}~\bibnamefont {{Bucher}}}, \bibinfo {author}
  {\bibfnamefont {C.}~\bibnamefont {{Burigana}}}, \bibinfo {author}
  {\bibfnamefont {E.}~\bibnamefont {{Calabrese}}}, \bibinfo {author}
  {\bibfnamefont {J.~F.}\ \bibnamefont {{Cardoso}}}, \bibinfo {author}
  {\bibfnamefont {J.}~\bibnamefont {{Carron}}}, \bibinfo {author}
  {\bibfnamefont {A.}~\bibnamefont {{Challinor}}}, \bibinfo {author}
  {\bibfnamefont {H.~C.}\ \bibnamefont {{Chiang}}}, \bibinfo {author}
  {\bibfnamefont {L.~P.~L.}\ \bibnamefont {{Colombo}}}, \bibinfo {author}
  {\bibfnamefont {C.}~\bibnamefont {{Combet}}}, \bibinfo {author}
  {\bibfnamefont {B.~P.}\ \bibnamefont {{Crill}}}, \bibinfo {author}
  {\bibfnamefont {F.}~\bibnamefont {{Cuttaia}}}, \bibinfo {author}
  {\bibfnamefont {P.}~\bibnamefont {{de Bernardis}}}, \bibinfo {author}
  {\bibfnamefont {G.}~\bibnamefont {{de Zotti}}}, \bibinfo {author}
  {\bibfnamefont {J.}~\bibnamefont {{Delabrouille}}}, \bibinfo {author}
  {\bibfnamefont {E.}~\bibnamefont {{Di Valentino}}}, \bibinfo {author}
  {\bibfnamefont {J.~M.}\ \bibnamefont {{Diego}}}, \bibinfo {author}
  {\bibfnamefont {O.}~\bibnamefont {{Dor{\'e}}}}, \bibinfo {author}
  {\bibfnamefont {M.}~\bibnamefont {{Douspis}}}, \bibinfo {author}
  {\bibfnamefont {A.}~\bibnamefont {{Ducout}}}, \bibinfo {author}
  {\bibfnamefont {X.}~\bibnamefont {{Dupac}}}, \bibinfo {author} {\bibfnamefont
  {G.}~\bibnamefont {{Efstathiou}}}, \bibinfo {author} {\bibfnamefont
  {F.}~\bibnamefont {{Elsner}}}, \bibinfo {author} {\bibfnamefont {T.~A.}\
  \bibnamefont {{En{\ss}lin}}}, \bibinfo {author} {\bibfnamefont {H.~K.}\
  \bibnamefont {{Eriksen}}}, \bibinfo {author} {\bibfnamefont {Y.}~\bibnamefont
  {{Fantaye}}}, \bibinfo {author} {\bibfnamefont {R.}~\bibnamefont
  {{Fernandez-Cobos}}}, \bibinfo {author} {\bibfnamefont {F.}~\bibnamefont
  {{Forastieri}}}, \bibinfo {author} {\bibfnamefont {M.}~\bibnamefont
  {{Frailis}}}, \bibinfo {author} {\bibfnamefont {A.~A.}\ \bibnamefont
  {{Fraisse}}}, \bibinfo {author} {\bibfnamefont {E.}~\bibnamefont
  {{Franceschi}}}, \bibinfo {author} {\bibfnamefont {A.}~\bibnamefont
  {{Frolov}}}, \bibinfo {author} {\bibfnamefont {S.}~\bibnamefont
  {{Galeotta}}}, \bibinfo {author} {\bibfnamefont {S.}~\bibnamefont {{Galli}}},
  \bibinfo {author} {\bibfnamefont {K.}~\bibnamefont {{Ganga}}}, \bibinfo
  {author} {\bibfnamefont {R.~T.}\ \bibnamefont {{G{\'e}nova-Santos}}},
  \bibinfo {author} {\bibfnamefont {M.}~\bibnamefont {{Gerbino}}}, \bibinfo
  {author} {\bibfnamefont {T.}~\bibnamefont {{Ghosh}}}, \bibinfo {author}
  {\bibfnamefont {J.}~\bibnamefont {{Gonz{\'a}lez-Nuevo}}}, \bibinfo {author}
  {\bibfnamefont {K.~M.}\ \bibnamefont {{G{\'o}rski}}}, \bibinfo {author}
  {\bibfnamefont {S.}~\bibnamefont {{Gratton}}}, \bibinfo {author}
  {\bibfnamefont {A.}~\bibnamefont {{Gruppuso}}}, \bibinfo {author}
  {\bibfnamefont {J.~E.}\ \bibnamefont {{Gudmundsson}}}, \bibinfo {author}
  {\bibfnamefont {J.}~\bibnamefont {{Hamann}}}, \bibinfo {author}
  {\bibfnamefont {W.}~\bibnamefont {{Hand ley}}}, \bibinfo {author}
  {\bibfnamefont {F.~K.}\ \bibnamefont {{Hansen}}}, \bibinfo {author}
  {\bibfnamefont {D.}~\bibnamefont {{Herranz}}}, \bibinfo {author}
  {\bibfnamefont {E.}~\bibnamefont {{Hivon}}}, \bibinfo {author} {\bibfnamefont
  {Z.}~\bibnamefont {{Huang}}}, \bibinfo {author} {\bibfnamefont {A.~H.}\
  \bibnamefont {{Jaffe}}}, \bibinfo {author} {\bibfnamefont {W.~C.}\
  \bibnamefont {{Jones}}}, \bibinfo {author} {\bibfnamefont {A.}~\bibnamefont
  {{Karakci}}}, \bibinfo {author} {\bibfnamefont {E.}~\bibnamefont
  {{Keih{\"a}nen}}}, \bibinfo {author} {\bibfnamefont {R.}~\bibnamefont
  {{Keskitalo}}}, \bibinfo {author} {\bibfnamefont {K.}~\bibnamefont
  {{Kiiveri}}}, \bibinfo {author} {\bibfnamefont {J.}~\bibnamefont {{Kim}}},
  \bibinfo {author} {\bibfnamefont {L.}~\bibnamefont {{Knox}}}, \bibinfo
  {author} {\bibfnamefont {N.}~\bibnamefont {{Krachmalnicoff}}}, \bibinfo
  {author} {\bibfnamefont {M.}~\bibnamefont {{Kunz}}}, \bibinfo {author}
  {\bibfnamefont {H.}~\bibnamefont {{Kurki-Suonio}}}, \bibinfo {author}
  {\bibfnamefont {G.}~\bibnamefont {{Lagache}}}, \bibinfo {author}
  {\bibfnamefont {J.~M.}\ \bibnamefont {{Lamarre}}}, \bibinfo {author}
  {\bibfnamefont {A.}~\bibnamefont {{Lasenby}}}, \bibinfo {author}
  {\bibfnamefont {M.}~\bibnamefont {{Lattanzi}}}, \bibinfo {author}
  {\bibfnamefont {C.~R.}\ \bibnamefont {{Lawrence}}}, \bibinfo {author}
  {\bibfnamefont {M.}~\bibnamefont {{Le Jeune}}}, \bibinfo {author}
  {\bibfnamefont {F.}~\bibnamefont {{Levrier}}}, \bibinfo {author}
  {\bibfnamefont {A.}~\bibnamefont {{Lewis}}}, \bibinfo {author} {\bibfnamefont
  {M.}~\bibnamefont {{Liguori}}}, \bibinfo {author} {\bibfnamefont {P.~B.}\
  \bibnamefont {{Lilje}}}, \bibinfo {author} {\bibfnamefont {V.}~\bibnamefont
  {{Lindholm}}}, \bibinfo {author} {\bibfnamefont {M.}~\bibnamefont
  {{L{\'o}pez-Caniego}}}, \bibinfo {author} {\bibfnamefont {P.~M.}\
  \bibnamefont {{Lubin}}}, \bibinfo {author} {\bibfnamefont {Y.~Z.}\
  \bibnamefont {{Ma}}}, \bibinfo {author} {\bibfnamefont {J.~F.}\ \bibnamefont
  {{Mac{\'\i}as-P{\'e}rez}}}, \bibinfo {author} {\bibfnamefont
  {G.}~\bibnamefont {{Maggio}}}, \bibinfo {author} {\bibfnamefont
  {D.}~\bibnamefont {{Maino}}}, \bibinfo {author} {\bibfnamefont
  {N.}~\bibnamefont {{Mandolesi}}}, \bibinfo {author} {\bibfnamefont
  {A.}~\bibnamefont {{Mangilli}}}, \bibinfo {author} {\bibfnamefont
  {A.}~\bibnamefont {{Marcos-Caballero}}}, \bibinfo {author} {\bibfnamefont
  {M.}~\bibnamefont {{Maris}}}, \bibinfo {author} {\bibfnamefont {P.~G.}\
  \bibnamefont {{Martin}}}, \bibinfo {author} {\bibfnamefont {E.}~\bibnamefont
  {{Mart{\'\i}nez-Gonz{\'a}lez}}}, \bibinfo {author} {\bibfnamefont
  {S.}~\bibnamefont {{Matarrese}}}, \bibinfo {author} {\bibfnamefont
  {N.}~\bibnamefont {{Mauri}}}, \bibinfo {author} {\bibfnamefont {J.~D.}\
  \bibnamefont {{McEwen}}}, \bibinfo {author} {\bibfnamefont {A.}~\bibnamefont
  {{Melchiorri}}}, \bibinfo {author} {\bibfnamefont {A.}~\bibnamefont
  {{Mennella}}}, \bibinfo {author} {\bibfnamefont {M.}~\bibnamefont
  {{Migliaccio}}}, \bibinfo {author} {\bibfnamefont {M.~A.}\ \bibnamefont
  {{Miville-Desch{\^e}nes}}}, \bibinfo {author} {\bibfnamefont
  {D.}~\bibnamefont {{Molinari}}}, \bibinfo {author} {\bibfnamefont
  {A.}~\bibnamefont {{Moneti}}}, \bibinfo {author} {\bibfnamefont
  {L.}~\bibnamefont {{Montier}}}, \bibinfo {author} {\bibfnamefont
  {G.}~\bibnamefont {{Morgante}}}, \bibinfo {author} {\bibfnamefont
  {A.}~\bibnamefont {{Moss}}}, \bibinfo {author} {\bibfnamefont
  {P.}~\bibnamefont {{Natoli}}}, \bibinfo {author} {\bibfnamefont
  {L.}~\bibnamefont {{Pagano}}}, \bibinfo {author} {\bibfnamefont
  {D.}~\bibnamefont {{Paoletti}}}, \bibinfo {author} {\bibfnamefont
  {B.}~\bibnamefont {{Partridge}}}, \bibinfo {author} {\bibfnamefont
  {G.}~\bibnamefont {{Patanchon}}}, \bibinfo {author} {\bibfnamefont
  {F.}~\bibnamefont {{Perrotta}}}, \bibinfo {author} {\bibfnamefont
  {V.}~\bibnamefont {{Pettorino}}}, \bibinfo {author} {\bibfnamefont
  {F.}~\bibnamefont {{Piacentini}}}, \bibinfo {author} {\bibfnamefont
  {L.}~\bibnamefont {{Polastri}}}, \bibinfo {author} {\bibfnamefont
  {G.}~\bibnamefont {{Polenta}}}, \bibinfo {author} {\bibfnamefont {J.~L.}\
  \bibnamefont {{Puget}}}, \bibinfo {author} {\bibfnamefont {J.~P.}\
  \bibnamefont {{Rachen}}}, \bibinfo {author} {\bibfnamefont {M.}~\bibnamefont
  {{Reinecke}}}, \bibinfo {author} {\bibfnamefont {M.}~\bibnamefont
  {{Remazeilles}}}, \bibinfo {author} {\bibfnamefont {A.}~\bibnamefont
  {{Renzi}}}, \bibinfo {author} {\bibfnamefont {G.}~\bibnamefont {{Rocha}}},
  \bibinfo {author} {\bibfnamefont {C.}~\bibnamefont {{Rosset}}}, \bibinfo
  {author} {\bibfnamefont {G.}~\bibnamefont {{Roudier}}}, \bibinfo {author}
  {\bibfnamefont {J.~A.}\ \bibnamefont {{Rubi{\~n}o-Mart{\'\i}n}}}, \bibinfo
  {author} {\bibfnamefont {B.}~\bibnamefont {{Ruiz-Granados}}}, \bibinfo
  {author} {\bibfnamefont {L.}~\bibnamefont {{Salvati}}}, \bibinfo {author}
  {\bibfnamefont {M.}~\bibnamefont {{Sandri}}}, \bibinfo {author}
  {\bibfnamefont {M.}~\bibnamefont {{Savelainen}}}, \bibinfo {author}
  {\bibfnamefont {D.}~\bibnamefont {{Scott}}}, \bibinfo {author} {\bibfnamefont
  {C.}~\bibnamefont {{Sirignano}}}, \bibinfo {author} {\bibfnamefont
  {R.}~\bibnamefont {{Sunyaev}}}, \bibinfo {author} {\bibfnamefont {A.~S.}\
  \bibnamefont {{Suur-Uski}}}, \bibinfo {author} {\bibfnamefont {J.~A.}\
  \bibnamefont {{Tauber}}}, \bibinfo {author} {\bibfnamefont {D.}~\bibnamefont
  {{Tavagnacco}}}, \bibinfo {author} {\bibfnamefont {M.}~\bibnamefont
  {{Tenti}}}, \bibinfo {author} {\bibfnamefont {L.}~\bibnamefont
  {{Toffolatti}}}, \bibinfo {author} {\bibfnamefont {M.}~\bibnamefont
  {{Tomasi}}}, \bibinfo {author} {\bibfnamefont {T.}~\bibnamefont
  {{Trombetti}}}, \bibinfo {author} {\bibfnamefont {J.}~\bibnamefont
  {{Valiviita}}}, \bibinfo {author} {\bibfnamefont {B.}~\bibnamefont {{Van
  Tent}}}, \bibinfo {author} {\bibfnamefont {P.}~\bibnamefont {{Vielva}}},
  \bibinfo {author} {\bibfnamefont {F.}~\bibnamefont {{Villa}}}, \bibinfo
  {author} {\bibfnamefont {N.}~\bibnamefont {{Vittorio}}}, \bibinfo {author}
  {\bibfnamefont {B.~D.}\ \bibnamefont {{Wandelt}}}, \bibinfo {author}
  {\bibfnamefont {I.~K.}\ \bibnamefont {{Wehus}}}, \bibinfo {author}
  {\bibfnamefont {M.}~\bibnamefont {{White}}}, \bibinfo {author} {\bibfnamefont
  {S.~D.~M.}\ \bibnamefont {{White}}}, \bibinfo {author} {\bibfnamefont
  {A.}~\bibnamefont {{Zacchei}}}, \ and\ \bibinfo {author} {\bibfnamefont
  {A.}~\bibnamefont {{Zonca}}},\ }\href@noop {} {\bibfield  {journal} {\bibinfo
   {journal} {arXiv e-prints}\ ,\ \bibinfo {eid} {arXiv:1807.06210}} (\bibinfo
  {year} {2018})},\ \Eprint {http://arxiv.org/abs/1807.06210} {arXiv:1807.06210
  [astro-ph.CO]} \BibitemShut {NoStop}%
\bibitem [{\citenamefont {Smith}\ \emph {et~al.}(2007)\citenamefont {Smith},
  \citenamefont {Zahn},\ and\ \citenamefont {Dore}}]{Smith:2007rg}%
  \BibitemOpen
  \bibfield  {author} {\bibinfo {author} {\bibfnamefont {K.~M.}\ \bibnamefont
  {Smith}}, \bibinfo {author} {\bibfnamefont {O.}~\bibnamefont {Zahn}}, \ and\
  \bibinfo {author} {\bibfnamefont {O.}~\bibnamefont {Dore}},\ }\href {\doibase
  10.1103/PhysRevD.76.043510} {\bibfield  {journal} {\bibinfo  {journal} {Phys.
  Rev.}\ }\textbf {\bibinfo {volume} {D76}},\ \bibinfo {pages} {043510}
  (\bibinfo {year} {2007})},\ \Eprint {http://arxiv.org/abs/0705.3980}
  {arXiv:0705.3980 [astro-ph]} \BibitemShut {NoStop}%
\bibitem [{\citenamefont {{Das}}\ \emph {et~al.}(2011)\citenamefont {{Das}},
  \citenamefont {{Sherwin}}, \citenamefont {{Aguirre}}, \citenamefont
  {{Appel}}, \citenamefont {{Bond}}, \citenamefont {{Carvalho}}, \citenamefont
  {{Devlin}}, \citenamefont {{Dunkley}}, \citenamefont {{D{\"u}nner}},
  \citenamefont {{Essinger-Hileman}}, \citenamefont {{Fowler}}, \citenamefont
  {{Hajian}}, \citenamefont {{Halpern}}, \citenamefont {{Hasselfield}},
  \citenamefont {{Hincks}}, \citenamefont {{Hlozek}}, \citenamefont
  {{Huffenberger}}, \citenamefont {{Hughes}}, \citenamefont {{Irwin}},
  \citenamefont {{Klein}}, \citenamefont {{Kosowsky}}, \citenamefont
  {{Lupton}}, \citenamefont {{Marriage}}, \citenamefont {{Marsden}},
  \citenamefont {{Menanteau}}, \citenamefont {{Moodley}}, \citenamefont
  {{Niemack}}, \citenamefont {{Nolta}}, \citenamefont {{Page}}, \citenamefont
  {{Parker}}, \citenamefont {{Reese}}, \citenamefont {{Schmitt}}, \citenamefont
  {{Sehgal}}, \citenamefont {{Sievers}}, \citenamefont {{Spergel}},
  \citenamefont {{Staggs}}, \citenamefont {{Swetz}}, \citenamefont {{Switzer}},
  \citenamefont {{Thornton}}, \citenamefont {{Visnjic}},\ and\ \citenamefont
  {{Wollack}}}]{2011PhRvL.107b1301D}%
  \BibitemOpen
  \bibfield  {author} {\bibinfo {author} {\bibfnamefont {S.}~\bibnamefont
  {{Das}}}, \bibinfo {author} {\bibfnamefont {B.~D.}\ \bibnamefont
  {{Sherwin}}}, \bibinfo {author} {\bibfnamefont {P.}~\bibnamefont
  {{Aguirre}}}, \bibinfo {author} {\bibfnamefont {J.~W.}\ \bibnamefont
  {{Appel}}}, \bibinfo {author} {\bibfnamefont {J.~R.}\ \bibnamefont {{Bond}}},
  \bibinfo {author} {\bibfnamefont {C.~S.}\ \bibnamefont {{Carvalho}}},
  \bibinfo {author} {\bibfnamefont {M.~J.}\ \bibnamefont {{Devlin}}}, \bibinfo
  {author} {\bibfnamefont {J.}~\bibnamefont {{Dunkley}}}, \bibinfo {author}
  {\bibfnamefont {R.}~\bibnamefont {{D{\"u}nner}}}, \bibinfo {author}
  {\bibfnamefont {T.}~\bibnamefont {{Essinger-Hileman}}}, \bibinfo {author}
  {\bibfnamefont {J.~W.}\ \bibnamefont {{Fowler}}}, \bibinfo {author}
  {\bibfnamefont {A.}~\bibnamefont {{Hajian}}}, \bibinfo {author}
  {\bibfnamefont {M.}~\bibnamefont {{Halpern}}}, \bibinfo {author}
  {\bibfnamefont {M.}~\bibnamefont {{Hasselfield}}}, \bibinfo {author}
  {\bibfnamefont {A.~D.}\ \bibnamefont {{Hincks}}}, \bibinfo {author}
  {\bibfnamefont {R.}~\bibnamefont {{Hlozek}}}, \bibinfo {author}
  {\bibfnamefont {K.~M.}\ \bibnamefont {{Huffenberger}}}, \bibinfo {author}
  {\bibfnamefont {J.~P.}\ \bibnamefont {{Hughes}}}, \bibinfo {author}
  {\bibfnamefont {K.~D.}\ \bibnamefont {{Irwin}}}, \bibinfo {author}
  {\bibfnamefont {J.}~\bibnamefont {{Klein}}}, \bibinfo {author} {\bibfnamefont
  {A.}~\bibnamefont {{Kosowsky}}}, \bibinfo {author} {\bibfnamefont {R.~H.}\
  \bibnamefont {{Lupton}}}, \bibinfo {author} {\bibfnamefont {T.~A.}\
  \bibnamefont {{Marriage}}}, \bibinfo {author} {\bibfnamefont
  {D.}~\bibnamefont {{Marsden}}}, \bibinfo {author} {\bibfnamefont
  {F.}~\bibnamefont {{Menanteau}}}, \bibinfo {author} {\bibfnamefont
  {K.}~\bibnamefont {{Moodley}}}, \bibinfo {author} {\bibfnamefont {M.~D.}\
  \bibnamefont {{Niemack}}}, \bibinfo {author} {\bibfnamefont {M.~R.}\
  \bibnamefont {{Nolta}}}, \bibinfo {author} {\bibfnamefont {L.~A.}\
  \bibnamefont {{Page}}}, \bibinfo {author} {\bibfnamefont {L.}~\bibnamefont
  {{Parker}}}, \bibinfo {author} {\bibfnamefont {E.~D.}\ \bibnamefont
  {{Reese}}}, \bibinfo {author} {\bibfnamefont {B.~L.}\ \bibnamefont
  {{Schmitt}}}, \bibinfo {author} {\bibfnamefont {N.}~\bibnamefont {{Sehgal}}},
  \bibinfo {author} {\bibfnamefont {J.}~\bibnamefont {{Sievers}}}, \bibinfo
  {author} {\bibfnamefont {D.~N.}\ \bibnamefont {{Spergel}}}, \bibinfo {author}
  {\bibfnamefont {S.~T.}\ \bibnamefont {{Staggs}}}, \bibinfo {author}
  {\bibfnamefont {D.~S.}\ \bibnamefont {{Swetz}}}, \bibinfo {author}
  {\bibfnamefont {E.~R.}\ \bibnamefont {{Switzer}}}, \bibinfo {author}
  {\bibfnamefont {R.}~\bibnamefont {{Thornton}}}, \bibinfo {author}
  {\bibfnamefont {K.}~\bibnamefont {{Visnjic}}}, \ and\ \bibinfo {author}
  {\bibfnamefont {E.}~\bibnamefont {{Wollack}}},\ }\href {\doibase
  10.1103/PhysRevLett.107.021301} {\bibfield  {journal} {\bibinfo  {journal}
  {Physical Review Letters}\ }\textbf {\bibinfo {volume} {107}},\ \bibinfo
  {eid} {021301} (\bibinfo {year} {2011})},\ \Eprint
  {http://arxiv.org/abs/1103.2124} {arXiv:1103.2124} \BibitemShut {NoStop}%
\bibitem [{\citenamefont {{van Engelen}}\ \emph {et~al.}(2012)\citenamefont
  {{van Engelen}}, \citenamefont {{Keisler}}, \citenamefont {{Zahn}},
  \citenamefont {{Aird}}, \citenamefont {{Benson}}, \citenamefont {{Bleem}},
  \citenamefont {{Carlstrom}}, \citenamefont {{Chang}}, \citenamefont {{Cho}},
  \citenamefont {{Crawford}}, \citenamefont {{Crites}}, \citenamefont {{de
  Haan}}, \citenamefont {{Dobbs}}, \citenamefont {{Dudley}}, \citenamefont
  {{George}}, \citenamefont {{Halverson}}, \citenamefont {{Holder}},
  \citenamefont {{Holzapfel}}, \citenamefont {{Hoover}}, \citenamefont {{Hou}},
  \citenamefont {{Hrubes}}, \citenamefont {{Joy}}, \citenamefont {{Knox}},
  \citenamefont {{Lee}}, \citenamefont {{Leitch}}, \citenamefont {{Lueker}},
  \citenamefont {{Luong-Van}}, \citenamefont {{McMahon}}, \citenamefont
  {{Mehl}}, \citenamefont {{Meyer}}, \citenamefont {{Millea}}, \citenamefont
  {{Mohr}}, \citenamefont {{Montroy}}, \citenamefont {{Natoli}}, \citenamefont
  {{Padin}}, \citenamefont {{Plagge}}, \citenamefont {{Pryke}}, \citenamefont
  {{Reichardt}}, \citenamefont {{Ruhl}}, \citenamefont {{Sayre}}, \citenamefont
  {{Schaffer}}, \citenamefont {{Shaw}}, \citenamefont {{Shirokoff}},
  \citenamefont {{Spieler}}, \citenamefont {{Staniszewski}}, \citenamefont
  {{Stark}}, \citenamefont {{Story}}, \citenamefont {{Vanderlinde}},
  \citenamefont {{Vieira}},\ and\ \citenamefont
  {{Williamson}}}]{2012ApJ...756..142V}%
  \BibitemOpen
  \bibfield  {author} {\bibinfo {author} {\bibfnamefont {A.}~\bibnamefont {{van
  Engelen}}}, \bibinfo {author} {\bibfnamefont {R.}~\bibnamefont {{Keisler}}},
  \bibinfo {author} {\bibfnamefont {O.}~\bibnamefont {{Zahn}}}, \bibinfo
  {author} {\bibfnamefont {K.~A.}\ \bibnamefont {{Aird}}}, \bibinfo {author}
  {\bibfnamefont {B.~A.}\ \bibnamefont {{Benson}}}, \bibinfo {author}
  {\bibfnamefont {L.~E.}\ \bibnamefont {{Bleem}}}, \bibinfo {author}
  {\bibfnamefont {J.~E.}\ \bibnamefont {{Carlstrom}}}, \bibinfo {author}
  {\bibfnamefont {C.~L.}\ \bibnamefont {{Chang}}}, \bibinfo {author}
  {\bibfnamefont {H.~M.}\ \bibnamefont {{Cho}}}, \bibinfo {author}
  {\bibfnamefont {T.~M.}\ \bibnamefont {{Crawford}}}, \bibinfo {author}
  {\bibfnamefont {A.~T.}\ \bibnamefont {{Crites}}}, \bibinfo {author}
  {\bibfnamefont {T.}~\bibnamefont {{de Haan}}}, \bibinfo {author}
  {\bibfnamefont {M.~A.}\ \bibnamefont {{Dobbs}}}, \bibinfo {author}
  {\bibfnamefont {J.}~\bibnamefont {{Dudley}}}, \bibinfo {author}
  {\bibfnamefont {E.~M.}\ \bibnamefont {{George}}}, \bibinfo {author}
  {\bibfnamefont {N.~W.}\ \bibnamefont {{Halverson}}}, \bibinfo {author}
  {\bibfnamefont {G.~P.}\ \bibnamefont {{Holder}}}, \bibinfo {author}
  {\bibfnamefont {W.~L.}\ \bibnamefont {{Holzapfel}}}, \bibinfo {author}
  {\bibfnamefont {S.}~\bibnamefont {{Hoover}}}, \bibinfo {author}
  {\bibfnamefont {Z.}~\bibnamefont {{Hou}}}, \bibinfo {author} {\bibfnamefont
  {J.~D.}\ \bibnamefont {{Hrubes}}}, \bibinfo {author} {\bibfnamefont
  {M.}~\bibnamefont {{Joy}}}, \bibinfo {author} {\bibfnamefont
  {L.}~\bibnamefont {{Knox}}}, \bibinfo {author} {\bibfnamefont {A.~T.}\
  \bibnamefont {{Lee}}}, \bibinfo {author} {\bibfnamefont {E.~M.}\ \bibnamefont
  {{Leitch}}}, \bibinfo {author} {\bibfnamefont {M.}~\bibnamefont {{Lueker}}},
  \bibinfo {author} {\bibfnamefont {D.}~\bibnamefont {{Luong-Van}}}, \bibinfo
  {author} {\bibfnamefont {J.~J.}\ \bibnamefont {{McMahon}}}, \bibinfo {author}
  {\bibfnamefont {J.}~\bibnamefont {{Mehl}}}, \bibinfo {author} {\bibfnamefont
  {S.~S.}\ \bibnamefont {{Meyer}}}, \bibinfo {author} {\bibfnamefont
  {M.}~\bibnamefont {{Millea}}}, \bibinfo {author} {\bibfnamefont {J.~J.}\
  \bibnamefont {{Mohr}}}, \bibinfo {author} {\bibfnamefont {T.~E.}\
  \bibnamefont {{Montroy}}}, \bibinfo {author} {\bibfnamefont {T.}~\bibnamefont
  {{Natoli}}}, \bibinfo {author} {\bibfnamefont {S.}~\bibnamefont {{Padin}}},
  \bibinfo {author} {\bibfnamefont {T.}~\bibnamefont {{Plagge}}}, \bibinfo
  {author} {\bibfnamefont {C.}~\bibnamefont {{Pryke}}}, \bibinfo {author}
  {\bibfnamefont {C.~L.}\ \bibnamefont {{Reichardt}}}, \bibinfo {author}
  {\bibfnamefont {J.~E.}\ \bibnamefont {{Ruhl}}}, \bibinfo {author}
  {\bibfnamefont {J.~T.}\ \bibnamefont {{Sayre}}}, \bibinfo {author}
  {\bibfnamefont {K.~K.}\ \bibnamefont {{Schaffer}}}, \bibinfo {author}
  {\bibfnamefont {L.}~\bibnamefont {{Shaw}}}, \bibinfo {author} {\bibfnamefont
  {E.}~\bibnamefont {{Shirokoff}}}, \bibinfo {author} {\bibfnamefont {H.~G.}\
  \bibnamefont {{Spieler}}}, \bibinfo {author} {\bibfnamefont {Z.}~\bibnamefont
  {{Staniszewski}}}, \bibinfo {author} {\bibfnamefont {A.~A.}\ \bibnamefont
  {{Stark}}}, \bibinfo {author} {\bibfnamefont {K.}~\bibnamefont {{Story}}},
  \bibinfo {author} {\bibfnamefont {K.}~\bibnamefont {{Vanderlinde}}}, \bibinfo
  {author} {\bibfnamefont {J.~D.}\ \bibnamefont {{Vieira}}}, \ and\ \bibinfo
  {author} {\bibfnamefont {R.}~\bibnamefont {{Williamson}}},\ }\href {\doibase
  10.1088/0004-637X/756/2/142} {\bibfield  {journal} {\bibinfo  {journal}
  {\apj}\ }\textbf {\bibinfo {volume} {756}},\ \bibinfo {eid} {142} (\bibinfo
  {year} {2012})},\ \Eprint {http://arxiv.org/abs/1202.0546} {arXiv:1202.0546
  [astro-ph.CO]} \BibitemShut {NoStop}%
\bibitem [{\citenamefont {{Omori}}\ \emph {et~al.}(2017)\citenamefont
  {{Omori}}, \citenamefont {{Chown}}, \citenamefont {{Simard}}, \citenamefont
  {{Story}}, \citenamefont {{Aylor}}, \citenamefont {{Baxter}}, \citenamefont
  {{Benson}}, \citenamefont {{Bleem}}, \citenamefont {{Carlstrom}},
  \citenamefont {{Chang}}, \citenamefont {{Cho}}, \citenamefont {{Crawford}},
  \citenamefont {{Crites}}, \citenamefont {{de Haan}}, \citenamefont {{Dobbs}},
  \citenamefont {{Everett}}, \citenamefont {{George}}, \citenamefont
  {{Halverson}}, \citenamefont {{Harrington}}, \citenamefont {{Holder}},
  \citenamefont {{Hou}}, \citenamefont {{Holzapfel}}, \citenamefont {{Hrubes}},
  \citenamefont {{Knox}}, \citenamefont {{Lee}}, \citenamefont {{Leitch}},
  \citenamefont {{Luong-Van}}, \citenamefont {{Manzotti}}, \citenamefont
  {{Marrone}}, \citenamefont {{McMahon}}, \citenamefont {{Meyer}},
  \citenamefont {{Mocanu}}, \citenamefont {{Mohr}}, \citenamefont {{Natoli}},
  \citenamefont {{Padin}}, \citenamefont {{Pryke}}, \citenamefont
  {{Reichardt}}, \citenamefont {{Ruhl}}, \citenamefont {{Sayre}}, \citenamefont
  {{Schaffer}}, \citenamefont {{Shirokoff}}, \citenamefont {{Staniszewski}},
  \citenamefont {{Stark}}, \citenamefont {{Vanderlinde}}, \citenamefont
  {{Vieira}}, \citenamefont {{Williamson}},\ and\ \citenamefont
  {{Zahn}}}]{2017ApJ...849..124O}%
  \BibitemOpen
  \bibfield  {author} {\bibinfo {author} {\bibfnamefont {Y.}~\bibnamefont
  {{Omori}}}, \bibinfo {author} {\bibfnamefont {R.}~\bibnamefont {{Chown}}},
  \bibinfo {author} {\bibfnamefont {G.}~\bibnamefont {{Simard}}}, \bibinfo
  {author} {\bibfnamefont {K.~T.}\ \bibnamefont {{Story}}}, \bibinfo {author}
  {\bibfnamefont {K.}~\bibnamefont {{Aylor}}}, \bibinfo {author} {\bibfnamefont
  {E.~J.}\ \bibnamefont {{Baxter}}}, \bibinfo {author} {\bibfnamefont {B.~A.}\
  \bibnamefont {{Benson}}}, \bibinfo {author} {\bibfnamefont {L.~E.}\
  \bibnamefont {{Bleem}}}, \bibinfo {author} {\bibfnamefont {J.~E.}\
  \bibnamefont {{Carlstrom}}}, \bibinfo {author} {\bibfnamefont {C.~L.}\
  \bibnamefont {{Chang}}}, \bibinfo {author} {\bibfnamefont {H.~M.}\
  \bibnamefont {{Cho}}}, \bibinfo {author} {\bibfnamefont {T.~M.}\ \bibnamefont
  {{Crawford}}}, \bibinfo {author} {\bibfnamefont {A.~T.}\ \bibnamefont
  {{Crites}}}, \bibinfo {author} {\bibfnamefont {T.}~\bibnamefont {{de Haan}}},
  \bibinfo {author} {\bibfnamefont {M.~A.}\ \bibnamefont {{Dobbs}}}, \bibinfo
  {author} {\bibfnamefont {W.~B.}\ \bibnamefont {{Everett}}}, \bibinfo {author}
  {\bibfnamefont {E.~M.}\ \bibnamefont {{George}}}, \bibinfo {author}
  {\bibfnamefont {N.~W.}\ \bibnamefont {{Halverson}}}, \bibinfo {author}
  {\bibfnamefont {N.~L.}\ \bibnamefont {{Harrington}}}, \bibinfo {author}
  {\bibfnamefont {G.~P.}\ \bibnamefont {{Holder}}}, \bibinfo {author}
  {\bibfnamefont {Z.}~\bibnamefont {{Hou}}}, \bibinfo {author} {\bibfnamefont
  {W.~L.}\ \bibnamefont {{Holzapfel}}}, \bibinfo {author} {\bibfnamefont
  {J.~D.}\ \bibnamefont {{Hrubes}}}, \bibinfo {author} {\bibfnamefont
  {L.}~\bibnamefont {{Knox}}}, \bibinfo {author} {\bibfnamefont {A.~T.}\
  \bibnamefont {{Lee}}}, \bibinfo {author} {\bibfnamefont {E.~M.}\ \bibnamefont
  {{Leitch}}}, \bibinfo {author} {\bibfnamefont {D.}~\bibnamefont
  {{Luong-Van}}}, \bibinfo {author} {\bibfnamefont {A.}~\bibnamefont
  {{Manzotti}}}, \bibinfo {author} {\bibfnamefont {D.~P.}\ \bibnamefont
  {{Marrone}}}, \bibinfo {author} {\bibfnamefont {J.~J.}\ \bibnamefont
  {{McMahon}}}, \bibinfo {author} {\bibfnamefont {S.~S.}\ \bibnamefont
  {{Meyer}}}, \bibinfo {author} {\bibfnamefont {L.~M.}\ \bibnamefont
  {{Mocanu}}}, \bibinfo {author} {\bibfnamefont {J.~J.}\ \bibnamefont
  {{Mohr}}}, \bibinfo {author} {\bibfnamefont {T.}~\bibnamefont {{Natoli}}},
  \bibinfo {author} {\bibfnamefont {S.}~\bibnamefont {{Padin}}}, \bibinfo
  {author} {\bibfnamefont {C.}~\bibnamefont {{Pryke}}}, \bibinfo {author}
  {\bibfnamefont {C.~L.}\ \bibnamefont {{Reichardt}}}, \bibinfo {author}
  {\bibfnamefont {J.~E.}\ \bibnamefont {{Ruhl}}}, \bibinfo {author}
  {\bibfnamefont {J.~T.}\ \bibnamefont {{Sayre}}}, \bibinfo {author}
  {\bibfnamefont {K.~K.}\ \bibnamefont {{Schaffer}}}, \bibinfo {author}
  {\bibfnamefont {E.}~\bibnamefont {{Shirokoff}}}, \bibinfo {author}
  {\bibfnamefont {Z.}~\bibnamefont {{Staniszewski}}}, \bibinfo {author}
  {\bibfnamefont {A.~A.}\ \bibnamefont {{Stark}}}, \bibinfo {author}
  {\bibfnamefont {K.}~\bibnamefont {{Vanderlinde}}}, \bibinfo {author}
  {\bibfnamefont {J.~D.}\ \bibnamefont {{Vieira}}}, \bibinfo {author}
  {\bibfnamefont {R.}~\bibnamefont {{Williamson}}}, \ and\ \bibinfo {author}
  {\bibfnamefont {O.}~\bibnamefont {{Zahn}}},\ }\href {\doibase
  10.3847/1538-4357/aa8d1d} {\bibfield  {journal} {\bibinfo  {journal} {\apj}\
  }\textbf {\bibinfo {volume} {849}},\ \bibinfo {eid} {124} (\bibinfo {year}
  {2017})},\ \Eprint {http://arxiv.org/abs/1705.00743} {arXiv:1705.00743
  [astro-ph.CO]} \BibitemShut {NoStop}%
\bibitem [{\citenamefont {Ade}\ \emph {et~al.}(2016)\citenamefont {Ade} \emph
  {et~al.}}]{Ade:2015zua}%
  \BibitemOpen
  \bibfield  {author} {\bibinfo {author} {\bibfnamefont {P.~A.~R.}\
  \bibnamefont {Ade}} \emph {et~al.} (\bibinfo {collaboration} {Planck}),\
  }\href {\doibase 10.1051/0004-6361/201525941} {\bibfield  {journal} {\bibinfo
   {journal} {Astron. Astrophys.}\ }\textbf {\bibinfo {volume} {594}},\
  \bibinfo {pages} {A15} (\bibinfo {year} {2016})},\ \Eprint
  {http://arxiv.org/abs/1502.01591} {arXiv:1502.01591 [astro-ph.CO]}
  \BibitemShut {NoStop}%
\bibitem [{\citenamefont {{Story}}\ \emph {et~al.}(2015)\citenamefont
  {{Story}}, \citenamefont {{Hanson}}, \citenamefont {{Ade}}, \citenamefont
  {{Aird}}, \citenamefont {{Austermann}}, \citenamefont {{Beall}},
  \citenamefont {{Bender}}, \citenamefont {{Benson}}, \citenamefont {{Bleem}},
  \citenamefont {{Carlstrom}}, \citenamefont {{Chang}}, \citenamefont
  {{Chiang}}, \citenamefont {{Cho}}, \citenamefont {{Citron}}, \citenamefont
  {{Crawford}}, \citenamefont {{Crites}}, \citenamefont {{de Haan}},
  \citenamefont {{Dobbs}}, \citenamefont {{Everett}}, \citenamefont
  {{Gallicchio}}, \citenamefont {{Gao}}, \citenamefont {{George}},
  \citenamefont {{Gilbert}}, \citenamefont {{Halverson}}, \citenamefont
  {{Harrington}}, \citenamefont {{Henning}}, \citenamefont {{Hilton}},
  \citenamefont {{Holder}}, \citenamefont {{Holzapfel}}, \citenamefont
  {{Hoover}}, \citenamefont {{Hou}}, \citenamefont {{Hrubes}}, \citenamefont
  {{Huang}}, \citenamefont {{Hubmayr}}, \citenamefont {{Irwin}}, \citenamefont
  {{Keisler}}, \citenamefont {{Knox}}, \citenamefont {{Lee}}, \citenamefont
  {{Leitch}}, \citenamefont {{Li}}, \citenamefont {{Liang}}, \citenamefont
  {{Luong-Van}}, \citenamefont {{McMahon}}, \citenamefont {{Mehl}},
  \citenamefont {{Meyer}}, \citenamefont {{Mocanu}}, \citenamefont {{Montroy}},
  \citenamefont {{Natoli}}, \citenamefont {{Nibarger}}, \citenamefont
  {{Novosad}}, \citenamefont {{Padin}}, \citenamefont {{Pryke}}, \citenamefont
  {{Reichardt}}, \citenamefont {{Ruhl}}, \citenamefont {{Saliwanchik}},
  \citenamefont {{Sayre}}, \citenamefont {{Schaffer}}, \citenamefont
  {{Smecher}}, \citenamefont {{Stark}}, \citenamefont {{Tucker}}, \citenamefont
  {{Vand erlinde}}, \citenamefont {{Vieira}}, \citenamefont {{Wang}},
  \citenamefont {{Whitehorn}}, \citenamefont {{Yefremenko}},\ and\
  \citenamefont {{Zahn}}}]{2015ApJ...810...50S}%
  \BibitemOpen
  \bibfield  {author} {\bibinfo {author} {\bibfnamefont {K.~T.}\ \bibnamefont
  {{Story}}}, \bibinfo {author} {\bibfnamefont {D.}~\bibnamefont {{Hanson}}},
  \bibinfo {author} {\bibfnamefont {P.~A.~R.}\ \bibnamefont {{Ade}}}, \bibinfo
  {author} {\bibfnamefont {K.~A.}\ \bibnamefont {{Aird}}}, \bibinfo {author}
  {\bibfnamefont {J.~E.}\ \bibnamefont {{Austermann}}}, \bibinfo {author}
  {\bibfnamefont {J.~A.}\ \bibnamefont {{Beall}}}, \bibinfo {author}
  {\bibfnamefont {A.~N.}\ \bibnamefont {{Bender}}}, \bibinfo {author}
  {\bibfnamefont {B.~A.}\ \bibnamefont {{Benson}}}, \bibinfo {author}
  {\bibfnamefont {L.~E.}\ \bibnamefont {{Bleem}}}, \bibinfo {author}
  {\bibfnamefont {J.~E.}\ \bibnamefont {{Carlstrom}}}, \bibinfo {author}
  {\bibfnamefont {C.~L.}\ \bibnamefont {{Chang}}}, \bibinfo {author}
  {\bibfnamefont {H.~C.}\ \bibnamefont {{Chiang}}}, \bibinfo {author}
  {\bibfnamefont {H.~M.}\ \bibnamefont {{Cho}}}, \bibinfo {author}
  {\bibfnamefont {R.}~\bibnamefont {{Citron}}}, \bibinfo {author}
  {\bibfnamefont {T.~M.}\ \bibnamefont {{Crawford}}}, \bibinfo {author}
  {\bibfnamefont {A.~T.}\ \bibnamefont {{Crites}}}, \bibinfo {author}
  {\bibfnamefont {T.}~\bibnamefont {{de Haan}}}, \bibinfo {author}
  {\bibfnamefont {M.~A.}\ \bibnamefont {{Dobbs}}}, \bibinfo {author}
  {\bibfnamefont {W.}~\bibnamefont {{Everett}}}, \bibinfo {author}
  {\bibfnamefont {J.}~\bibnamefont {{Gallicchio}}}, \bibinfo {author}
  {\bibfnamefont {J.}~\bibnamefont {{Gao}}}, \bibinfo {author} {\bibfnamefont
  {E.~M.}\ \bibnamefont {{George}}}, \bibinfo {author} {\bibfnamefont
  {A.}~\bibnamefont {{Gilbert}}}, \bibinfo {author} {\bibfnamefont {N.~W.}\
  \bibnamefont {{Halverson}}}, \bibinfo {author} {\bibfnamefont
  {N.}~\bibnamefont {{Harrington}}}, \bibinfo {author} {\bibfnamefont {J.~W.}\
  \bibnamefont {{Henning}}}, \bibinfo {author} {\bibfnamefont {G.~C.}\
  \bibnamefont {{Hilton}}}, \bibinfo {author} {\bibfnamefont {G.~P.}\
  \bibnamefont {{Holder}}}, \bibinfo {author} {\bibfnamefont {W.~L.}\
  \bibnamefont {{Holzapfel}}}, \bibinfo {author} {\bibfnamefont
  {S.}~\bibnamefont {{Hoover}}}, \bibinfo {author} {\bibfnamefont
  {Z.}~\bibnamefont {{Hou}}}, \bibinfo {author} {\bibfnamefont {J.~D.}\
  \bibnamefont {{Hrubes}}}, \bibinfo {author} {\bibfnamefont {N.}~\bibnamefont
  {{Huang}}}, \bibinfo {author} {\bibfnamefont {J.}~\bibnamefont {{Hubmayr}}},
  \bibinfo {author} {\bibfnamefont {K.~D.}\ \bibnamefont {{Irwin}}}, \bibinfo
  {author} {\bibfnamefont {R.}~\bibnamefont {{Keisler}}}, \bibinfo {author}
  {\bibfnamefont {L.}~\bibnamefont {{Knox}}}, \bibinfo {author} {\bibfnamefont
  {A.~T.}\ \bibnamefont {{Lee}}}, \bibinfo {author} {\bibfnamefont {E.~M.}\
  \bibnamefont {{Leitch}}}, \bibinfo {author} {\bibfnamefont {D.}~\bibnamefont
  {{Li}}}, \bibinfo {author} {\bibfnamefont {C.}~\bibnamefont {{Liang}}},
  \bibinfo {author} {\bibfnamefont {D.}~\bibnamefont {{Luong-Van}}}, \bibinfo
  {author} {\bibfnamefont {J.~J.}\ \bibnamefont {{McMahon}}}, \bibinfo {author}
  {\bibfnamefont {J.}~\bibnamefont {{Mehl}}}, \bibinfo {author} {\bibfnamefont
  {S.~S.}\ \bibnamefont {{Meyer}}}, \bibinfo {author} {\bibfnamefont
  {L.}~\bibnamefont {{Mocanu}}}, \bibinfo {author} {\bibfnamefont {T.~E.}\
  \bibnamefont {{Montroy}}}, \bibinfo {author} {\bibfnamefont {T.}~\bibnamefont
  {{Natoli}}}, \bibinfo {author} {\bibfnamefont {J.~P.}\ \bibnamefont
  {{Nibarger}}}, \bibinfo {author} {\bibfnamefont {V.}~\bibnamefont
  {{Novosad}}}, \bibinfo {author} {\bibfnamefont {S.}~\bibnamefont {{Padin}}},
  \bibinfo {author} {\bibfnamefont {C.}~\bibnamefont {{Pryke}}}, \bibinfo
  {author} {\bibfnamefont {C.~L.}\ \bibnamefont {{Reichardt}}}, \bibinfo
  {author} {\bibfnamefont {J.~E.}\ \bibnamefont {{Ruhl}}}, \bibinfo {author}
  {\bibfnamefont {B.~R.}\ \bibnamefont {{Saliwanchik}}}, \bibinfo {author}
  {\bibfnamefont {J.~T.}\ \bibnamefont {{Sayre}}}, \bibinfo {author}
  {\bibfnamefont {K.~K.}\ \bibnamefont {{Schaffer}}}, \bibinfo {author}
  {\bibfnamefont {G.}~\bibnamefont {{Smecher}}}, \bibinfo {author}
  {\bibfnamefont {A.~A.}\ \bibnamefont {{Stark}}}, \bibinfo {author}
  {\bibfnamefont {C.}~\bibnamefont {{Tucker}}}, \bibinfo {author}
  {\bibfnamefont {K.}~\bibnamefont {{Vand erlinde}}}, \bibinfo {author}
  {\bibfnamefont {J.~D.}\ \bibnamefont {{Vieira}}}, \bibinfo {author}
  {\bibfnamefont {G.}~\bibnamefont {{Wang}}}, \bibinfo {author} {\bibfnamefont
  {N.}~\bibnamefont {{Whitehorn}}}, \bibinfo {author} {\bibfnamefont
  {V.}~\bibnamefont {{Yefremenko}}}, \ and\ \bibinfo {author} {\bibfnamefont
  {O.}~\bibnamefont {{Zahn}}},\ }\href {\doibase 10.1088/0004-637X/810/1/50}
  {\bibfield  {journal} {\bibinfo  {journal} {\apj}\ }\textbf {\bibinfo
  {volume} {810}},\ \bibinfo {eid} {50} (\bibinfo {year} {2015})},\ \Eprint
  {http://arxiv.org/abs/1412.4760} {arXiv:1412.4760 [astro-ph.CO]} \BibitemShut
  {NoStop}%
\bibitem [{\citenamefont {Ade}\ \emph {et~al.}(2014)\citenamefont {Ade} \emph
  {et~al.}}]{Ade:2013gez}%
  \BibitemOpen
  \bibfield  {author} {\bibinfo {author} {\bibfnamefont {P.~A.~R.}\
  \bibnamefont {Ade}} \emph {et~al.} (\bibinfo {collaboration} {POLARBEAR}),\
  }\href {\doibase 10.1103/PhysRevLett.113.021301} {\bibfield  {journal}
  {\bibinfo  {journal} {Phys. Rev. Lett.}\ }\textbf {\bibinfo {volume} {113}},\
  \bibinfo {pages} {021301} (\bibinfo {year} {2014})},\ \Eprint
  {http://arxiv.org/abs/1312.6646} {arXiv:1312.6646 [astro-ph.CO]} \BibitemShut
  {NoStop}%
\bibitem [{\citenamefont {Hu}\ and\ \citenamefont
  {Okamoto}(2002)}]{hu2002mass}%
  \BibitemOpen
  \bibfield  {author} {\bibinfo {author} {\bibfnamefont {W.}~\bibnamefont
  {Hu}}\ and\ \bibinfo {author} {\bibfnamefont {T.}~\bibnamefont {Okamoto}},\
  }\href {\doibase 10.1086/341110} {\bibfield  {journal} {\bibinfo  {journal}
  {Astrophys. J.}\ }\textbf {\bibinfo {volume} {574}},\ \bibinfo {pages} {566}
  (\bibinfo {year} {2002})},\ \Eprint {http://arxiv.org/abs/astro-ph/0111606}
  {arXiv:astro-ph/0111606 [astro-ph]} \BibitemShut {NoStop}%
\bibitem [{\citenamefont {Hirata}\ and\ \citenamefont
  {Seljak}(2003)}]{hirata2003analyzing}%
  \BibitemOpen
  \bibfield  {author} {\bibinfo {author} {\bibfnamefont {C.~M.}\ \bibnamefont
  {Hirata}}\ and\ \bibinfo {author} {\bibfnamefont {U.}~\bibnamefont
  {Seljak}},\ }\href {\doibase 10.1103/PhysRevD.67.043001} {\bibfield
  {journal} {\bibinfo  {journal} {Phys. Rev.}\ }\textbf {\bibinfo {volume}
  {D67}},\ \bibinfo {pages} {043001} (\bibinfo {year} {2003})},\ \Eprint
  {http://arxiv.org/abs/astro-ph/0209489} {arXiv:astro-ph/0209489 [astro-ph]}
  \BibitemShut {NoStop}%
\bibitem [{\citenamefont {{Horowitz}}\ \emph {et~al.}(2017)\citenamefont
  {{Horowitz}}, \citenamefont {{Ferraro}},\ and\ \citenamefont
  {{Sherwin}}}]{Horowitz:2017iql}%
  \BibitemOpen
  \bibfield  {author} {\bibinfo {author} {\bibfnamefont {B.}~\bibnamefont
  {{Horowitz}}}, \bibinfo {author} {\bibfnamefont {S.}~\bibnamefont
  {{Ferraro}}}, \ and\ \bibinfo {author} {\bibfnamefont {B.~D.}\ \bibnamefont
  {{Sherwin}}},\ }\href@noop {} {\bibfield  {journal} {\bibinfo  {journal}
  {ArXiv e-prints}\ } (\bibinfo {year} {2017})},\ \Eprint
  {http://arxiv.org/abs/1710.10236} {arXiv:1710.10236} \BibitemShut {NoStop}%
\bibitem [{\citenamefont {Carron}\ and\ \citenamefont
  {Lewis}(2017)}]{carron2017maximum}%
  \BibitemOpen
  \bibfield  {author} {\bibinfo {author} {\bibfnamefont {J.}~\bibnamefont
  {Carron}}\ and\ \bibinfo {author} {\bibfnamefont {A.}~\bibnamefont {Lewis}},\
  }\href {\doibase 10.1103/PhysRevD.96.063510} {\bibfield  {journal} {\bibinfo
  {journal} {Phys. Rev.}\ }\textbf {\bibinfo {volume} {D96}},\ \bibinfo {pages}
  {063510} (\bibinfo {year} {2017})},\ \Eprint
  {http://arxiv.org/abs/1704.08230} {arXiv:1704.08230 [astro-ph.CO]}
  \BibitemShut {NoStop}%
\bibitem [{\citenamefont {Seljak}\ and\ \citenamefont
  {Zaldarriaga}(2000)}]{seljak2000lensing}%
  \BibitemOpen
  \bibfield  {author} {\bibinfo {author} {\bibfnamefont {U.}~\bibnamefont
  {Seljak}}\ and\ \bibinfo {author} {\bibfnamefont {M.}~\bibnamefont
  {Zaldarriaga}},\ }\href {\doibase 10.1086/309098} {\bibfield  {journal}
  {\bibinfo  {journal} {Astrophys. J.}\ }\textbf {\bibinfo {volume} {538}},\
  \bibinfo {pages} {57} (\bibinfo {year} {2000})},\ \Eprint
  {http://arxiv.org/abs/astro-ph/9907254} {arXiv:astro-ph/9907254 [astro-ph]}
  \BibitemShut {NoStop}%
\bibitem [{\citenamefont {Hu}\ \emph {et~al.}(2007{\natexlab{a}})\citenamefont
  {Hu}, \citenamefont {DeDeo},\ and\ \citenamefont {Vale}}]{hu2007cluster}%
  \BibitemOpen
  \bibfield  {author} {\bibinfo {author} {\bibfnamefont {W.}~\bibnamefont
  {Hu}}, \bibinfo {author} {\bibfnamefont {S.}~\bibnamefont {DeDeo}}, \ and\
  \bibinfo {author} {\bibfnamefont {C.}~\bibnamefont {Vale}},\ }\href {\doibase
  10.1088/1367-2630/9/12/441} {\bibfield  {journal} {\bibinfo  {journal} {New
  J. Phys.}\ }\textbf {\bibinfo {volume} {9}},\ \bibinfo {pages} {441}
  (\bibinfo {year} {2007}{\natexlab{a}})},\ \Eprint
  {http://arxiv.org/abs/astro-ph/0701276} {arXiv:astro-ph/0701276 [astro-ph]}
  \BibitemShut {NoStop}%
\bibitem [{\citenamefont {{Efstathiou}}(2004)}]{2004MNRAS.349..603E}%
  \BibitemOpen
  \bibfield  {author} {\bibinfo {author} {\bibfnamefont {G.}~\bibnamefont
  {{Efstathiou}}},\ }\href {\doibase 10.1111/j.1365-2966.2004.07530.x}
  {\bibfield  {journal} {\bibinfo  {journal} {\mnras}\ }\textbf {\bibinfo
  {volume} {349}},\ \bibinfo {pages} {603} (\bibinfo {year} {2004})},\ \Eprint
  {http://arxiv.org/abs/astro-ph/0307515} {arXiv:astro-ph/0307515 [astro-ph]}
  \BibitemShut {NoStop}%
\bibitem [{\citenamefont {Nguyen}\ \emph {et~al.}(2019)\citenamefont {Nguyen},
  \citenamefont {Sehgal},\ and\ \citenamefont
  {Madhavacheril}}]{Nguyen:2017zqu}%
  \BibitemOpen
  \bibfield  {author} {\bibinfo {author} {\bibfnamefont {H.~N.}\ \bibnamefont
  {Nguyen}}, \bibinfo {author} {\bibfnamefont {N.}~\bibnamefont {Sehgal}}, \
  and\ \bibinfo {author} {\bibfnamefont {M.}~\bibnamefont {Madhavacheril}},\
  }\href {\doibase 10.1103/PhysRevD.99.023502} {\bibfield  {journal} {\bibinfo
  {journal} {Phys. Rev.}\ }\textbf {\bibinfo {volume} {D99}},\ \bibinfo {pages}
  {023502} (\bibinfo {year} {2019})},\ \Eprint
  {http://arxiv.org/abs/1710.03747} {arXiv:1710.03747 [astro-ph.CO]}
  \BibitemShut {NoStop}%
\bibitem [{\citenamefont {Hu}\ \emph {et~al.}(2007{\natexlab{b}})\citenamefont
  {Hu}, \citenamefont {DeDeo},\ and\ \citenamefont {Vale}}]{Hu_2007}%
  \BibitemOpen
  \bibfield  {author} {\bibinfo {author} {\bibfnamefont {W.}~\bibnamefont
  {Hu}}, \bibinfo {author} {\bibfnamefont {S.}~\bibnamefont {DeDeo}}, \ and\
  \bibinfo {author} {\bibfnamefont {C.}~\bibnamefont {Vale}},\ }\href {\doibase
  10.1088/1367-2630/9/12/441} {\bibfield  {journal} {\bibinfo  {journal} {New
  Journal of Physics}\ }\textbf {\bibinfo {volume} {9}},\ \bibinfo {pages}
  {441} (\bibinfo {year} {2007}{\natexlab{b}})}\BibitemShut {NoStop}%
\bibitem [{\citenamefont {{Namikawa}}\ \emph {et~al.}(2013)\citenamefont
  {{Namikawa}}, \citenamefont {{Hanson}},\ and\ \citenamefont
  {{Takahashi}}}]{2013MNRAS.431..609N}%
  \BibitemOpen
  \bibfield  {author} {\bibinfo {author} {\bibfnamefont {T.}~\bibnamefont
  {{Namikawa}}}, \bibinfo {author} {\bibfnamefont {D.}~\bibnamefont
  {{Hanson}}}, \ and\ \bibinfo {author} {\bibfnamefont {R.}~\bibnamefont
  {{Takahashi}}},\ }\href {\doibase 10.1093/mnras/stt195} {\bibfield  {journal}
  {\bibinfo  {journal} {\mnras}\ }\textbf {\bibinfo {volume} {431}},\ \bibinfo
  {pages} {609} (\bibinfo {year} {2013})},\ \Eprint
  {http://arxiv.org/abs/1209.0091} {arXiv:1209.0091 [astro-ph.CO]} \BibitemShut
  {NoStop}%
\bibitem [{\citenamefont {Sherwin}\ \emph {et~al.}(2017)\citenamefont
  {Sherwin}, \citenamefont {van Engelen}, \citenamefont {Sehgal}, \citenamefont
  {Madhavacheril}, \citenamefont {Addison}, \citenamefont {Aiola},
  \citenamefont {Allison}, \citenamefont {Battaglia}, \citenamefont {Becker},
  \citenamefont {Beall}, \citenamefont {Bond}, \citenamefont {Calabrese},
  \citenamefont {Datta}, \citenamefont {Devlin}, \citenamefont {D\"unner},
  \citenamefont {Dunkley}, \citenamefont {Fox}, \citenamefont {Gallardo},
  \citenamefont {Halpern}, \citenamefont {Hasselfield}, \citenamefont
  {Henderson}, \citenamefont {Hill}, \citenamefont {Hilton}, \citenamefont
  {Hubmayr}, \citenamefont {Hughes}, \citenamefont {Hincks}, \citenamefont
  {Hlozek}, \citenamefont {Huffenberger}, \citenamefont {Koopman},
  \citenamefont {Kosowsky}, \citenamefont {Louis}, \citenamefont {Maurin},
  \citenamefont {McMahon}, \citenamefont {Moodley}, \citenamefont {Naess},
  \citenamefont {Nati}, \citenamefont {Newburgh}, \citenamefont {Niemack},
  \citenamefont {Page}, \citenamefont {Sievers}, \citenamefont {Spergel},
  \citenamefont {Staggs}, \citenamefont {Thornton}, \citenamefont {Van~Lanen},
  \citenamefont {Vavagiakis},\ and\ \citenamefont
  {Wollack}}]{PhysRevD.95.123529}%
  \BibitemOpen
  \bibfield  {author} {\bibinfo {author} {\bibfnamefont {B.~D.}\ \bibnamefont
  {Sherwin}}, \bibinfo {author} {\bibfnamefont {A.}~\bibnamefont {van
  Engelen}}, \bibinfo {author} {\bibfnamefont {N.}~\bibnamefont {Sehgal}},
  \bibinfo {author} {\bibfnamefont {M.}~\bibnamefont {Madhavacheril}}, \bibinfo
  {author} {\bibfnamefont {G.~E.}\ \bibnamefont {Addison}}, \bibinfo {author}
  {\bibfnamefont {S.}~\bibnamefont {Aiola}}, \bibinfo {author} {\bibfnamefont
  {R.}~\bibnamefont {Allison}}, \bibinfo {author} {\bibfnamefont
  {N.}~\bibnamefont {Battaglia}}, \bibinfo {author} {\bibfnamefont {D.~T.}\
  \bibnamefont {Becker}}, \bibinfo {author} {\bibfnamefont {J.~A.}\
  \bibnamefont {Beall}}, \bibinfo {author} {\bibfnamefont {J.~R.}\ \bibnamefont
  {Bond}}, \bibinfo {author} {\bibfnamefont {E.}~\bibnamefont {Calabrese}},
  \bibinfo {author} {\bibfnamefont {R.}~\bibnamefont {Datta}}, \bibinfo
  {author} {\bibfnamefont {M.~J.}\ \bibnamefont {Devlin}}, \bibinfo {author}
  {\bibfnamefont {R.}~\bibnamefont {D\"unner}}, \bibinfo {author}
  {\bibfnamefont {J.}~\bibnamefont {Dunkley}}, \bibinfo {author} {\bibfnamefont
  {A.~E.}\ \bibnamefont {Fox}}, \bibinfo {author} {\bibfnamefont
  {P.}~\bibnamefont {Gallardo}}, \bibinfo {author} {\bibfnamefont
  {M.}~\bibnamefont {Halpern}}, \bibinfo {author} {\bibfnamefont
  {M.}~\bibnamefont {Hasselfield}}, \bibinfo {author} {\bibfnamefont
  {S.}~\bibnamefont {Henderson}}, \bibinfo {author} {\bibfnamefont {J.~C.}\
  \bibnamefont {Hill}}, \bibinfo {author} {\bibfnamefont {G.~C.}\ \bibnamefont
  {Hilton}}, \bibinfo {author} {\bibfnamefont {J.}~\bibnamefont {Hubmayr}},
  \bibinfo {author} {\bibfnamefont {J.~P.}\ \bibnamefont {Hughes}}, \bibinfo
  {author} {\bibfnamefont {A.~D.}\ \bibnamefont {Hincks}}, \bibinfo {author}
  {\bibfnamefont {R.}~\bibnamefont {Hlozek}}, \bibinfo {author} {\bibfnamefont
  {K.~M.}\ \bibnamefont {Huffenberger}}, \bibinfo {author} {\bibfnamefont
  {B.}~\bibnamefont {Koopman}}, \bibinfo {author} {\bibfnamefont
  {A.}~\bibnamefont {Kosowsky}}, \bibinfo {author} {\bibfnamefont
  {T.}~\bibnamefont {Louis}}, \bibinfo {author} {\bibfnamefont
  {L.}~\bibnamefont {Maurin}}, \bibinfo {author} {\bibfnamefont
  {J.}~\bibnamefont {McMahon}}, \bibinfo {author} {\bibfnamefont
  {K.}~\bibnamefont {Moodley}}, \bibinfo {author} {\bibfnamefont
  {S.}~\bibnamefont {Naess}}, \bibinfo {author} {\bibfnamefont
  {F.}~\bibnamefont {Nati}}, \bibinfo {author} {\bibfnamefont {L.}~\bibnamefont
  {Newburgh}}, \bibinfo {author} {\bibfnamefont {M.~D.}\ \bibnamefont
  {Niemack}}, \bibinfo {author} {\bibfnamefont {L.~A.}\ \bibnamefont {Page}},
  \bibinfo {author} {\bibfnamefont {J.}~\bibnamefont {Sievers}}, \bibinfo
  {author} {\bibfnamefont {D.~N.}\ \bibnamefont {Spergel}}, \bibinfo {author}
  {\bibfnamefont {S.~T.}\ \bibnamefont {Staggs}}, \bibinfo {author}
  {\bibfnamefont {R.~J.}\ \bibnamefont {Thornton}}, \bibinfo {author}
  {\bibfnamefont {J.}~\bibnamefont {Van~Lanen}}, \bibinfo {author}
  {\bibfnamefont {E.}~\bibnamefont {Vavagiakis}}, \ and\ \bibinfo {author}
  {\bibfnamefont {E.~J.}\ \bibnamefont {Wollack}},\ }\href {\doibase
  10.1103/PhysRevD.95.123529} {\bibfield  {journal} {\bibinfo  {journal} {Phys.
  Rev. D}\ }\textbf {\bibinfo {volume} {95}},\ \bibinfo {pages} {123529}
  (\bibinfo {year} {2017})}\BibitemShut {NoStop}%
\bibitem [{\citenamefont {{Nguy{\^e}n}}\ \emph {et~al.}(2019)\citenamefont
  {{Nguy{\^e}n}}, \citenamefont {{Sehgal}},\ and\ \citenamefont
  {{Madhavacheril}}}]{2019PhRvD..99b3502N}%
  \BibitemOpen
  \bibfield  {author} {\bibinfo {author} {\bibfnamefont {H.~N.}\ \bibnamefont
  {{Nguy{\^e}n}}}, \bibinfo {author} {\bibfnamefont {N.}~\bibnamefont
  {{Sehgal}}}, \ and\ \bibinfo {author} {\bibfnamefont {M.~S.}\ \bibnamefont
  {{Madhavacheril}}},\ }\href {\doibase 10.1103/PhysRevD.99.023502} {\bibfield
  {journal} {\bibinfo  {journal} {\prd}\ }\textbf {\bibinfo {volume} {99}},\
  \bibinfo {eid} {023502} (\bibinfo {year} {2019})},\ \Eprint
  {http://arxiv.org/abs/1710.03747} {arXiv:1710.03747 [astro-ph.CO]}
  \BibitemShut {NoStop}%
\bibitem [{\citenamefont {Knox}(1995)}]{knox}%
  \BibitemOpen
  \bibfield  {author} {\bibinfo {author} {\bibfnamefont {L.}~\bibnamefont
  {Knox}},\ }\href {\doibase 10.1103/PhysRevD.52.4307} {\bibfield  {journal}
  {\bibinfo  {journal} {Phys. Rev.}\ }\textbf {\bibinfo {volume} {D52}},\
  \bibinfo {pages} {4307} (\bibinfo {year} {1995})},\ \Eprint
  {http://arxiv.org/abs/astro-ph/9504054} {arXiv:astro-ph/9504054 [astro-ph]}
  \BibitemShut {NoStop}%
\bibitem [{\citenamefont {Ferraro}\ and\ \citenamefont
  {Hill}(2018)}]{Ferraro:2017fac}%
  \BibitemOpen
  \bibfield  {author} {\bibinfo {author} {\bibfnamefont {S.}~\bibnamefont
  {Ferraro}}\ and\ \bibinfo {author} {\bibfnamefont {J.~C.}\ \bibnamefont
  {Hill}},\ }\href {\doibase 10.1103/PhysRevD.97.023512} {\bibfield  {journal}
  {\bibinfo  {journal} {Phys. Rev.}\ }\textbf {\bibinfo {volume} {D97}},\
  \bibinfo {pages} {023512} (\bibinfo {year} {2018})},\ \Eprint
  {http://arxiv.org/abs/1705.06751} {arXiv:1705.06751 [astro-ph.CO]}
  \BibitemShut {NoStop}%
\bibitem [{\citenamefont {Smith}\ and\ \citenamefont
  {Ferraro}(2017)}]{Smith:2016lnt}%
  \BibitemOpen
  \bibfield  {author} {\bibinfo {author} {\bibfnamefont {K.~M.}\ \bibnamefont
  {Smith}}\ and\ \bibinfo {author} {\bibfnamefont {S.}~\bibnamefont
  {Ferraro}},\ }\href {\doibase 10.1103/PhysRevLett.119.021301} {\bibfield
  {journal} {\bibinfo  {journal} {Phys. Rev. Lett.}\ }\textbf {\bibinfo
  {volume} {119}},\ \bibinfo {pages} {021301} (\bibinfo {year} {2017})},\
  \Eprint {http://arxiv.org/abs/1607.01769} {arXiv:1607.01769 [astro-ph.CO]}
  \BibitemShut {NoStop}%
\bibitem [{\citenamefont {Amblard}\ \emph {et~al.}(2004)\citenamefont
  {Amblard}, \citenamefont {Vale},\ and\ \citenamefont
  {White}}]{Amblard:2004ih}%
  \BibitemOpen
  \bibfield  {author} {\bibinfo {author} {\bibfnamefont {A.}~\bibnamefont
  {Amblard}}, \bibinfo {author} {\bibfnamefont {C.}~\bibnamefont {Vale}}, \
  and\ \bibinfo {author} {\bibfnamefont {M.~J.}\ \bibnamefont {White}},\ }\href
  {\doibase 10.1016/j.newast.2004.05.003} {\bibfield  {journal} {\bibinfo
  {journal} {New Astron.}\ }\textbf {\bibinfo {volume} {9}},\ \bibinfo {pages}
  {687} (\bibinfo {year} {2004})},\ \Eprint
  {http://arxiv.org/abs/astro-ph/0403075} {arXiv:astro-ph/0403075 [astro-ph]}
  \BibitemShut {NoStop}%
\bibitem [{\citenamefont {{Millea}}\ \emph {et~al.}(2017)\citenamefont
  {{Millea}}, \citenamefont {{Anderes}},\ and\ \citenamefont
  {{Wandelt}}}]{Millea:2017fyd}%
  \BibitemOpen
  \bibfield  {author} {\bibinfo {author} {\bibfnamefont {M.}~\bibnamefont
  {{Millea}}}, \bibinfo {author} {\bibfnamefont {E.}~\bibnamefont {{Anderes}}},
  \ and\ \bibinfo {author} {\bibfnamefont {B.~D.}\ \bibnamefont {{Wandelt}}},\
  }\href@noop {} {\bibfield  {journal} {\bibinfo  {journal} {ArXiv e-prints}\ }
  (\bibinfo {year} {2017})},\ \Eprint {http://arxiv.org/abs/1708.06753}
  {arXiv:1708.06753} \BibitemShut {NoStop}%
\bibitem [{\citenamefont {{Sehgal}}\ \emph {et~al.}(2019)\citenamefont
  {{Sehgal}}, \citenamefont {{Nguyen}}, \citenamefont {{Meyers}}, \citenamefont
  {{Munchmeyer}}, \citenamefont {{Mroczkowski}}, \citenamefont {{Di Mascolo}},
  \citenamefont {{Baxter}}, \citenamefont {{Cyr-Racine}}, \citenamefont
  {{Madhavacheril}}, \citenamefont {{Beringue}}, \citenamefont {{Holder}},
  \citenamefont {{Nagai}}, \citenamefont {{Dicker}}, \citenamefont {{Dvorkin}},
  \citenamefont {{Ferraro}}, \citenamefont {{Fuller}}, \citenamefont
  {{Gluscevic}}, \citenamefont {{Han}}, \citenamefont {{Jain}}, \citenamefont
  {{Johnson}}, \citenamefont {{Klaassen}}, \citenamefont {{Meerburg}},
  \citenamefont {{Motloch}}, \citenamefont {{Spergel}}, \citenamefont {{van
  Engelen}}, \citenamefont {{Adshead}}, \citenamefont {{Armstrong}},
  \citenamefont {{Baccigalupi}}, \citenamefont {{Barron}}, \citenamefont
  {{Basu}}, \citenamefont {{Benson}}, \citenamefont {{Beutler}}, \citenamefont
  {{Bond}}, \citenamefont {{Borrill}}, \citenamefont {{Calabrese}},
  \citenamefont {{Darwish}}, \citenamefont {{Denny}}, \citenamefont
  {{Douglass}}, \citenamefont {{Essinger-Hileman}}, \citenamefont {{Foreman}},
  \citenamefont {{Frayer}}, \citenamefont {{Gerbino}}, \citenamefont
  {{Gontcho}}, \citenamefont {{Grohs}}, \citenamefont {{Gupta}}, \citenamefont
  {{Hill}}, \citenamefont {{Hirata}}, \citenamefont {{Hotinli}}, \citenamefont
  {{Johnson}}, \citenamefont {{Kamionkowski}}, \citenamefont {{Kovetz}},
  \citenamefont {{Lau}}, \citenamefont {{Liguori}}, \citenamefont {{Namikawa}},
  \citenamefont {{Newburgh}}, \citenamefont {{Partridge}}, \citenamefont
  {{Piacentni}}, \citenamefont {{Rose}}, \citenamefont {{Rossi}}, \citenamefont
  {{Saliwanchik}}, \citenamefont {{Schaan}}, \citenamefont {{Shan}},
  \citenamefont {{Simon}}, \citenamefont {{Slosar}}, \citenamefont {{Switzer}},
  \citenamefont {{Trac}}, \citenamefont {{Xu}}, \citenamefont {{Zaldarriaga}},\
  and\ \citenamefont {{Zemcov}}}]{CMBHD}%
  \BibitemOpen
  \bibfield  {author} {\bibinfo {author} {\bibfnamefont {N.}~\bibnamefont
  {{Sehgal}}}, \bibinfo {author} {\bibfnamefont {H.~N.}\ \bibnamefont
  {{Nguyen}}}, \bibinfo {author} {\bibfnamefont {J.}~\bibnamefont {{Meyers}}},
  \bibinfo {author} {\bibfnamefont {M.}~\bibnamefont {{Munchmeyer}}}, \bibinfo
  {author} {\bibfnamefont {T.}~\bibnamefont {{Mroczkowski}}}, \bibinfo {author}
  {\bibfnamefont {L.}~\bibnamefont {{Di Mascolo}}}, \bibinfo {author}
  {\bibfnamefont {E.}~\bibnamefont {{Baxter}}}, \bibinfo {author}
  {\bibfnamefont {F.-Y.}\ \bibnamefont {{Cyr-Racine}}}, \bibinfo {author}
  {\bibfnamefont {M.}~\bibnamefont {{Madhavacheril}}}, \bibinfo {author}
  {\bibfnamefont {B.}~\bibnamefont {{Beringue}}}, \bibinfo {author}
  {\bibfnamefont {G.}~\bibnamefont {{Holder}}}, \bibinfo {author}
  {\bibfnamefont {D.}~\bibnamefont {{Nagai}}}, \bibinfo {author} {\bibfnamefont
  {S.}~\bibnamefont {{Dicker}}}, \bibinfo {author} {\bibfnamefont
  {C.}~\bibnamefont {{Dvorkin}}}, \bibinfo {author} {\bibfnamefont
  {S.}~\bibnamefont {{Ferraro}}}, \bibinfo {author} {\bibfnamefont {G.~M.}\
  \bibnamefont {{Fuller}}}, \bibinfo {author} {\bibfnamefont {V.}~\bibnamefont
  {{Gluscevic}}}, \bibinfo {author} {\bibfnamefont {D.}~\bibnamefont {{Han}}},
  \bibinfo {author} {\bibfnamefont {B.}~\bibnamefont {{Jain}}}, \bibinfo
  {author} {\bibfnamefont {B.}~\bibnamefont {{Johnson}}}, \bibinfo {author}
  {\bibfnamefont {P.}~\bibnamefont {{Klaassen}}}, \bibinfo {author}
  {\bibfnamefont {D.}~\bibnamefont {{Meerburg}}}, \bibinfo {author}
  {\bibfnamefont {P.}~\bibnamefont {{Motloch}}}, \bibinfo {author}
  {\bibfnamefont {D.~N.}\ \bibnamefont {{Spergel}}}, \bibinfo {author}
  {\bibfnamefont {A.}~\bibnamefont {{van Engelen}}}, \bibinfo {author}
  {\bibfnamefont {P.}~\bibnamefont {{Adshead}}}, \bibinfo {author}
  {\bibfnamefont {R.}~\bibnamefont {{Armstrong}}}, \bibinfo {author}
  {\bibfnamefont {C.}~\bibnamefont {{Baccigalupi}}}, \bibinfo {author}
  {\bibfnamefont {D.}~\bibnamefont {{Barron}}}, \bibinfo {author}
  {\bibfnamefont {K.}~\bibnamefont {{Basu}}}, \bibinfo {author} {\bibfnamefont
  {B.}~\bibnamefont {{Benson}}}, \bibinfo {author} {\bibfnamefont
  {F.}~\bibnamefont {{Beutler}}}, \bibinfo {author} {\bibfnamefont {J.~R.}\
  \bibnamefont {{Bond}}}, \bibinfo {author} {\bibfnamefont {J.}~\bibnamefont
  {{Borrill}}}, \bibinfo {author} {\bibfnamefont {E.}~\bibnamefont
  {{Calabrese}}}, \bibinfo {author} {\bibfnamefont {O.}~\bibnamefont
  {{Darwish}}}, \bibinfo {author} {\bibfnamefont {S.~L.}\ \bibnamefont
  {{Denny}}}, \bibinfo {author} {\bibfnamefont {K.~A.}\ \bibnamefont
  {{Douglass}}}, \bibinfo {author} {\bibfnamefont {T.}~\bibnamefont
  {{Essinger-Hileman}}}, \bibinfo {author} {\bibfnamefont {S.}~\bibnamefont
  {{Foreman}}}, \bibinfo {author} {\bibfnamefont {D.}~\bibnamefont {{Frayer}}},
  \bibinfo {author} {\bibfnamefont {M.}~\bibnamefont {{Gerbino}}}, \bibinfo
  {author} {\bibfnamefont {S.~G.~A.}\ \bibnamefont {{Gontcho}}}, \bibinfo
  {author} {\bibfnamefont {E.~B.}\ \bibnamefont {{Grohs}}}, \bibinfo {author}
  {\bibfnamefont {N.}~\bibnamefont {{Gupta}}}, \bibinfo {author} {\bibfnamefont
  {J.~C.}\ \bibnamefont {{Hill}}}, \bibinfo {author} {\bibfnamefont {C.~M.}\
  \bibnamefont {{Hirata}}}, \bibinfo {author} {\bibfnamefont {S.}~\bibnamefont
  {{Hotinli}}}, \bibinfo {author} {\bibfnamefont {M.~C.}\ \bibnamefont
  {{Johnson}}}, \bibinfo {author} {\bibfnamefont {M.}~\bibnamefont
  {{Kamionkowski}}}, \bibinfo {author} {\bibfnamefont {E.~D.}\ \bibnamefont
  {{Kovetz}}}, \bibinfo {author} {\bibfnamefont {E.~T.}\ \bibnamefont {{Lau}}},
  \bibinfo {author} {\bibfnamefont {M.}~\bibnamefont {{Liguori}}}, \bibinfo
  {author} {\bibfnamefont {T.}~\bibnamefont {{Namikawa}}}, \bibinfo {author}
  {\bibfnamefont {L.}~\bibnamefont {{Newburgh}}}, \bibinfo {author}
  {\bibfnamefont {B.}~\bibnamefont {{Partridge}}}, \bibinfo {author}
  {\bibfnamefont {F.}~\bibnamefont {{Piacentni}}}, \bibinfo {author}
  {\bibfnamefont {B.}~\bibnamefont {{Rose}}}, \bibinfo {author} {\bibfnamefont
  {G.}~\bibnamefont {{Rossi}}}, \bibinfo {author} {\bibfnamefont
  {B.}~\bibnamefont {{Saliwanchik}}}, \bibinfo {author} {\bibfnamefont
  {E.}~\bibnamefont {{Schaan}}}, \bibinfo {author} {\bibfnamefont
  {H.}~\bibnamefont {{Shan}}}, \bibinfo {author} {\bibfnamefont
  {S.}~\bibnamefont {{Simon}}}, \bibinfo {author} {\bibfnamefont
  {A.}~\bibnamefont {{Slosar}}}, \bibinfo {author} {\bibfnamefont {E.~R.}\
  \bibnamefont {{Switzer}}}, \bibinfo {author} {\bibfnamefont {H.}~\bibnamefont
  {{Trac}}}, \bibinfo {author} {\bibfnamefont {W.}~\bibnamefont {{Xu}}},
  \bibinfo {author} {\bibfnamefont {M.}~\bibnamefont {{Zaldarriaga}}}, \ and\
  \bibinfo {author} {\bibfnamefont {M.}~\bibnamefont {{Zemcov}}},\ }\href@noop
  {} {\bibfield  {journal} {\bibinfo  {journal} {arXiv e-prints}\ ,\ \bibinfo
  {eid} {arXiv:1903.03263}} (\bibinfo {year} {2019})},\ \Eprint
  {http://arxiv.org/abs/1903.03263} {arXiv:1903.03263 [astro-ph.CO]}
  \BibitemShut {NoStop}%
\bibitem [{\citenamefont {Lewis}\ and\ \citenamefont
  {Challinor}(2006)}]{Lewis:2006fu}%
  \BibitemOpen
  \bibfield  {author} {\bibinfo {author} {\bibfnamefont {A.}~\bibnamefont
  {Lewis}}\ and\ \bibinfo {author} {\bibfnamefont {A.}~\bibnamefont
  {Challinor}},\ }\href {\doibase 10.1016/j.physrep.2006.03.002} {\bibfield
  {journal} {\bibinfo  {journal} {Phys. Rept.}\ }\textbf {\bibinfo {volume}
  {429}},\ \bibinfo {pages} {1} (\bibinfo {year} {2006})},\ \Eprint
  {http://arxiv.org/abs/astro-ph/0601594} {arXiv:astro-ph/0601594 [astro-ph]}
  \BibitemShut {NoStop}%
\end{thebibliography}%

\end{document}